\documentclass[prl,aps,twocolumn,superscriptaddress, 10pt]{revtex4-2}

\usepackage[pdftex]{graphicx}
\usepackage{amsbsy, amssymb, amsmath, bm, mathtools}
\usepackage{xspace}
\usepackage{float}
\usepackage{bm}% bold math
\usepackage{color}
\usepackage{tabularx}
\usepackage{booktabs}
\usepackage[section]{placeins}
\usepackage[colorlinks=true,linkcolor=blue,citecolor=blue]{hyperref}
\usepackage{url}
\usepackage{threeparttable}
\usepackage{multirow}
\usepackage{enumerate}
\usepackage{enumitem}
\usepackage{color}
\usepackage{subfigure}  %added.
\usepackage{epstopdf}
\usepackage{bbm}
\usepackage{graphicx}
\usepackage{hyperref}
\usepackage{threeparttable}
\usepackage{amsthm}
\usepackage{mathtools}
\usepackage{color}
\usepackage{tikz}
\usepackage{float}
\usepackage{empheq}
\usepackage{algorithm}
\usepackage{algpseudocode}

\begin{document}
	
\title{Two-dimensional XY Ferromagnet Induced by Long-range Interaction}

\author{Tianning Xiao}
\thanks{These two authors contributed equally to this work}
\affiliation{Hefei National Research Center for Physical Sciences at the Microscale and School of Physical Sciences, University of Science and Technology of China, Hefei 230026, China}

\author{Dingyun Yao}
\thanks{These two authors contributed equally to this work}
\affiliation{Hefei National Research Center for Physical Sciences at the Microscale and School of Physical Sciences, University of Science and Technology of China, Hefei 230026, China}

\author{Chao Zhang}
\email{zhangchao1986sdu@gmail.com}
\affiliation{Hefei National Research Center for Physical Sciences at the Microscale and School of Physical Sciences, University of Science and Technology of China, Hefei 230026, China}
\affiliation{Department of Physics, Anhui Normal University, Wuhu, Anhui 241000, China}

\author{Zhijie Fan}
\email{zfanac@ustc.edu.cn}
\affiliation{Hefei National Research Center for Physical Sciences at the Microscale and School of Physical Sciences, University of Science and Technology of China, Hefei 230026, China}
\affiliation{Hefei National Laboratory, University of Science and Technology of China, Hefei 230088, China}
\affiliation{Shanghai Research Center for Quantum Science and CAS Center for Excellence in Quantum Information and Quantum Physics, University of Science and Technology of China, Shanghai 201315, China}

\author{Youjin Deng}
\email{yjdeng@ustc.edu.cn}
\affiliation{Hefei National Research Center for Physical Sciences at the Microscale and School of Physical Sciences, University of Science and Technology of China, Hefei 230026, China}
\affiliation{Hefei National Laboratory, University of Science and Technology of China, Hefei 230088, China}
\affiliation{Shanghai Research Center for Quantum Science and CAS Center for Excellence in Quantum Information and Quantum Physics, University of Science and Technology of China, Shanghai 201315, China}

\begin{abstract}
The crossover between short-range and long-range (LR) universal behaviors remains a central theme in the physics of long-range interacting systems. The competition between LR coupling and the Berezinskii-Kosterlitz-Thouless mechanism makes the problem more subtle and less understood in the two-dimensional (2D) XY model, a cornerstone for investigating low-dimensional phenomena and their implications in quantum computation. We study the 2D XY model with algebraically decaying interaction $\sim1/r^{2+\sigma}$. Utilizing an advanced update strategy, we conduct large-scale Monte Carlo simulations of the model up to a linear size of $L=8192$. Our results demonstrate continuous phase transitions into a ferromagnetic phase for $\sigma \leq 2$, which exhibits the simultaneous emergence of a long-ranged order and a power-law decaying correlation function due to the Goldstone mode. Furthermore, we find logarithmic scaling behaviors in the low-temperature phase at $\sigma = 2$. The observed scaling behaviors in the low-temperature phase for $\sigma \le 2$ agree with our theoretical analysis. Our findings request further theoretical understandings and can be of practical application in cutting-edge experiments like Rydberg atom arrays. 
\end{abstract}

\maketitle

\textit{Introduction}.--- Long-range (LR) interacting systems have been studied in statistical and condensed matter physics for decades, unveiling a range of exotic physical phenomena~\cite{spivak2004, lahaye2009, peter2012}. This interest has recently intensified, driven by the experimental realizations of such systems in atomic, molecular, and optical (AMO) setups~\cite{saffman2010, lu2012, schauss2012, firstenberg2013, yan2013, aikawa2012, lewis2023}. In particular, the two-dimensional (2D) XY model with LR interactions has gained notable attention~\cite{giachetti2021, giachetti2022, chen2023, chen2023a}. Without LR interactions, the model undergoes the celebrated BKT transition driven by topological defects~\cite{kosterlitz2017} and serves as a fundamental cornerstone for understanding low-dimensional superfluidity~\cite{bishop1978} and superconductivity~\cite{epstein1981, benfatto2007, fazio2001}. Upon incorporating LR interactions, however, it becomes a pivotal framework for exploring the complex interplay between LR interactions and the BKT mechanism~\cite{kosterlitz2017}. Most importantly, recent implementations of the model in trapped ion setups and the Rydberg systems demonstrate its significance in quantum computation~\cite{lewis2023, chen2023, chen2023a}. 

The XY model belongs to the classical O$(\mathcal{N})$ spin models with $\mathcal{N} = 2$. The $d$-dimensional LR O$(\mathcal{N})$ spin model with power-law decaying $\sim 1/r^{d+\sigma}$ interactions has been extensively investigated, particularly regarding the renormalization group (RG) relevance of the LR interactions~\cite{kunz1976, fisher1972,sak1973,defenu2023,angelini2014, brezin2014, defenu2017,defenu2015}. In such systems, there exists a threshold value $\sigma_*$ separating the LR and SR critical behaviors. For $\sigma > \sigma_*$, the system is in the same universality class as its nearest-neighbor (NN) counterpart, while for $\sigma \le \sigma_*$, the LR interactions become relevant, yielding distinct critical properties~\cite{fisher1972,sak1973,defenu2023}. The value of $\sigma_*$ was first obtained in the seminal paper of Fisher et al.~\cite{fisher1972}, where a second-order $\epsilon$-expansion approach suggests $\sigma_* = 2$. Later, a new threshold $\sigma_* = 2 - \eta_{\text{SR}}$ was proposed by Sak~\cite{sak1973}, currently known as Sak's criterion, where $\eta_{\text{SR}}$ is the anomalous dimension in SR limit. While several numerical studies seemingly support Sak's criterion~\cite{luijten2002,angelini2014,horita2017}, other investigation and theoretical analysis favor the $\sigma_*=2$ scenario~\cite{vanenter1982, grassberger2013, blanchard2013}.

The problem becomes more subtle for the 2D XY model. In the SR limit, the Mermin-Wagner theorem forbids the formation of a long-range-order (LRO) phase~\cite{mermin1966}. Yet, the model undergoes a BKT transition, entering a quasi-long-range-order (QLRO) phase~\cite{kosterlitz2017}. Applying Sak's criterion to the 2D XY model can be especially nuanced because, rather than a single fixed point, the SR critical behavior is governed by an entire line of fixed points with a temperature-dependent anomalous dimension $\eta(T)$, and the phase transition is of topological type~\cite{giachetti2021, giachetti2022, kosterlitz2017}. Conventional strategies for analyzing the XY model, such as mapping it to Coulomb gas or the sine-Gordon model~\cite{minnhagen1987,gulacsi1998}, might fail in the presence of LR interaction~\cite{giachetti2021}. Furthermore, the numerical study of this model faces considerable difficulties, including logarithmic corrections owing to BKT universality~\cite{chen2022, kosterlitz2017}, severe finite-size effects, and the escalating computational costs associated with LR interactions~\cite{luijten1995,michel2019}.

Recent field-theoretical studies of the 2D LR XY model predict an exotic phase diagram~\cite{giachetti2021,giachetti2022}. An intermediate QLRO phase is stabilized for $1.75 < \sigma < 2$, below which the system enters an LRO phase. Intriguingly, a similar study on the LR Villain model reveals different behavior~\cite{giachetti2023}, despite both models belonging to the same universality class in the SR limit~\cite{villain1975,jose1977}. This deviation is particularly notable given that such an intermediate QLRO phase is absent in previous numerical results of the LR diluted XY model in 2D~\cite{cescatti2019}, a model expected to share the same critical behaviors as the 2D LR XY model~\cite{berganza2013, cescatti2019}. 

\begin{figure}[t]
    \centering
    \includegraphics{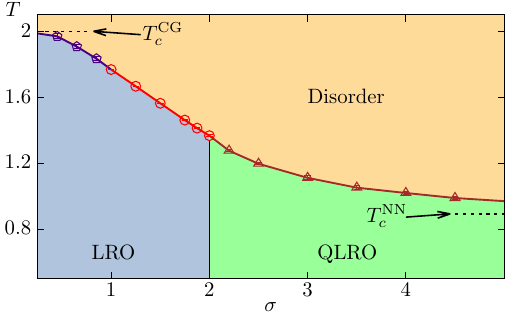}
    \vspace*{-6mm}
    \caption{Phase diagram of the long-range XY model in 2D. The SR regime ($\sigma > 2$) exhibits BKT transitions (brown line) into the QLRO phase. In the non-classical regime ($1 < \sigma \le 2$), the system undergoes a second-order transition (red line) into an LRO phase. Finally, in the classical regime ($\sigma \le 1$), the transition (purple lines) is described by the Gaussian theory. Symbol $T_c^{\text{CG}}$ stands for the critical temperatures for the complete-graph (CG) case and $T_c^{\text{NN}}$ for the nearest-neighbor (NN) case.}
    \label{fig_1_phase_diagram}
\end{figure}

In this Letter, we study the 2D LR XY model with power-law decaying $\sim 1/r^{d+\sigma}$ interactions by large-scale simulations up to a linear size of $L=8192$. The phase diagram of the model, as depicted in Fig.~\ref{fig_1_phase_diagram}, is characterized by three distinct regimes: the \textit{classical} regime ($\sigma \le 1$), the \textit{non-classical} regime ($1 < \sigma \leq 2$), and the \textit{short-range} regime ($\sigma > 2$). As expected, for $\sigma < 1$, the critical behaviors are governed by Gaussian mean-field theory~\cite{defenu2023}, while for $\sigma > 2$, the system exhibits BKT transitions. The non-classical regime ($1 < \sigma \leq 2$) is of particular interest. The finite-size scaling (FSS) behaviors in this regime demonstrate that the system undergoes a second-order transition with $\sigma$-dependent critical exponents. The ferromagnetic low-$T$ phase also features a power-law decaying spin correlation function $g(x) \sim g_0 + c x^{-\eta_\ell}$ originating from the Goldstone mode, with $g_0$ and $c$ being some constants and $\eta_\ell = 2 - \sigma$. Moreover, at $\sigma = 2$, we clearly observe that, as the criticality is gradually approached, the growing behavior of correlation length $\xi$ looks more and more different from that of the BKT transition and is well-described by a power-law behavior for a continuous phase transition. By exploring the FSS behaviors at a fixed temperature $T<T_c$ as a function of $\sigma$, we obtain another strong evidence that $\sigma=2$ is the threshold separating the LRO ferromagnet from the QLRO phase for $\sigma>2$.

\begin{figure*}[t]
    \centering
    \includegraphics{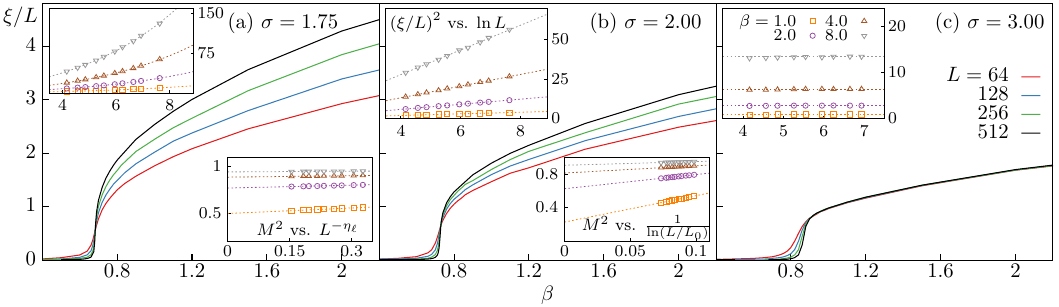}
    \vspace*{-6mm}
    \caption{Emergence of the LRO for $\sigma \leq 2$. As temperature $T$ decreases, the correlation-length ratio $\xi/L$ when $\sigma=1.75$ (a) and $2$ (b) displays typical scaling behaviors for a system entering into a long-range ordered phase via a continuous phase transition at $T_c$: it has an asymptotically universal value at $T=T_c$ and diverges for $T<T_c$ as $L$ increases. In contrast, for $\sigma=3$ (c), which has a BKT transition, $\xi/L$ for different $L$s quickly converges to a smooth function for $T<T_c$, as a consequence of the algebraically decaying QLRO.
    The top left insets further show that in the low-$T$ phase (with $\beta=1.0,2.0,4.0,8.0$), $(\xi/L)^2 \sim L^{\eta_\ell}$ with a $T$-independent exponent $\eta_\ell$ for $\sigma =1.75$ and $\sim \ln (L/L_0)$ for the marginal case $\sigma =2$, with $L_0 = 0.0025$ (the appearance of this logarithmic divergence is argued in the text).
    Moreover, the squared magnetization density $M^2$ for $T<T_c$ converges to non-zero constants as $L \to \infty$, providing direct evidence for the existence of long-range order.}
    \label{fig_2_T_scan}
\end{figure*}

\textit{Model, Algorithm and Observables}.---
Let us consider the LR interacting XY model on a square lattice of side length $L$,
\begin{align}
    \mathcal{H} = - \sum_{i < j} \frac{J}{r_{i,j}^{d+\sigma}} \bm{S}_i \cdot \bm{S}_j,
    \label{eq_Hamiltonian}
\end{align}
where $\bm{S}_{i}$ and $\bm{S}_j$ are $2$-component unit spin vectors at sites $i$ and $j$, respectively, and $r_{i,j}$ denotes the distance between these sites. The summation encompasses all unique pairs of spins. With periodic boundary conditions, each spin interacts with other $N-1$ spins ($N=L^2$) via the shortest distance. In addition, the interaction strength $J$ is normalized such that $\sum_{j>0} J/r_{0,j}^{2+\sigma} = 4$, to satisfy the strict extensitivity of the total energy and thus to reduce unnecessary finite-size corrections. ~\cite{filinov2010, horita2017, luijten1997}. The Boltzmann weight of a configuration is $\exp(-\beta \mathcal{H})$, with $\beta=1/k_{B}T$ the inverse temperature ($k_{B}=1$ is set).

Substantial computational expense is the primary factor hindering large-scale simulations of the model. In conventional Monte Carlo methods, it scales as $\cal{O}$($N$) per spin update due to LR interactions. Specialized techniques have been developed to efficiently simulate LR interacting systems~\cite{luijten1995,luijten1997,horita2017,michel2019}. We employ an enhanced version of the Luijten-Bl\"{o}te (LB) algorithm~\cite{luijten1995,luijten1997}, which utilizes cluster spin updates~\cite{swendsen1987,wolff1989} alongside an exceedingly efficient cluster construction procedure. This technique significantly accelerates the construction of clusters, rendering the computational time per spin independent of $N$. Specifically, we incorporate the clock sampling technique~\cite{michel2019} to efficiently sample bond activation events, substantially improving computational speed and memory usage. Also, it eliminates the need for a look-up table and alleviates truncation errors stemming from discrete cumulative probability integration approximations~\cite{luijten1997}.

Various physical quantities are measured. For a given configuration, we sample the magnetization density $M = L^{-2} \left| \sum_{i} \bm{S}_i \right|$ and its Fourier transform $M_k = L^{-2} \left| \sum_i \bm{S}_i e^{i\bm{k} \cdot \bm{r}_i} \right|$. Here, $\bm{r}_i$ denotes the coordinates of site $i$ and $\bm{k} = (2\pi/L, 0)$ is the smallest wave vector along the x-axis. We then obtain the susceptibility $\chi = L^2\langle M^2 \rangle$, the Fourier-transformed susceptibility $\chi_k = L^2 \langle M_k^2 \rangle$, where $\langle\cdot\rangle$ represents the statistical average. Finally, we define the second-moment correlation length $\xi = 1 /\left[2 \sin(|\bm{k}|/2)\right] \sqrt{\langle \chi \rangle/\langle \chi_k \rangle - 1}$~\cite{viet2009,ding2014,komura2012a,surungan2019}.

\textit{Results}.--- 
Dimensionless quantities, such as the Binder cumulant~\cite{binder1981} and the correlation-length ratio $\xi/L$~\cite{viet2009,ding2014,komura2012a,surungan2019}, are powerful tools in studying phase transitions. Figure~\ref{fig_2_T_scan} shows that for $\sigma \leq 2$, the $\xi/L$ curves display the typical FSS behaviors of a second-order transition, i.e., $\xi/L$ curves of different $L$s share a universal intersection point at $T = T_c$ and diverge for $T < T_c$ as $L$ increases~\cite{viet2009,ding2014,komura2012a,surungan2019}. The least-square fits, based on the standard FSS ansatz, successfully give an accurate estimation of critical points and critical exponents in the non-classical regime $1 \leq \sigma \leq 2$, as presented in Table~\ref{table_LRXY_fitting}. As a reference, characteristic FSS behavior of BKT transitions is observed for $\sigma = 3$, where $\xi/L$ curves converge to a non-trivial smooth function for $T < T_c$~\cite{tuan2022,kosterlitz2017}. These results suggest a threshold value $\sigma_* = 2$ in the LR XY model, below which the system develops a long-range order parameter and becomes a ferromagnet.

The spontaneous O($2$) symmetry breaking for $\sigma \le 2$ naturally implies the existence of Goldstone mode in the low-$T$ phase. Consider the field-theoretical Hamiltonian of 2D LR O$(\mathcal{N})$ models in momentum-space,
\begin{align}
    \beta H&=\int{\frac{\mathrm d^2q}{(2\pi)^2}(\frac{K_{2}}{2}q^2+K_{\sigma} q^\sigma)\Psi(\boldsymbol{q})\cdot\Psi(-\boldsymbol{q})} \\ \nonumber
    &+\int\mathrm{d}^2x(\frac t2\Psi^2+u\Psi^4). 
\end{align}
where $\Psi$ is the $\mathcal{N}$-component order parameter field, $t$ is the distance to criticality, and $K_2$, $K_\sigma$, $u$ are coupling constants. For $\sigma < 2$, $K_{\sigma}q^\sigma$ is the leading term, and thus $\frac {K_2}{2} q^2$ can be ignored. In the LRO phase where $t < 0$, employing the saddle point approximation, $\Psi$ can then be written in terms of longitudinal and transverse fluctuations $\Psi(\bm{x}) = \Psi_L(\bm{x}) + \Psi_T(\bm{x})$. In this expansion, the two-point correlation of transverse fluctuation in the momentum space $\langle\Psi_T(\bm{q})\Psi_T(-\bm{q})\rangle$ is proportional to $|\bm{q}|^{-\sigma}$, which results in a power-law correlation in real-space $\langle\Psi_T(0)\Psi_T(\boldsymbol{x})\rangle \sim |\bm{x}|^{-2 + \sigma}$. Therefore, for the LR XY model, the correlation function in the LRO phase is then, $g(x) =g_0+ c x^{-\eta_\ell}$, where $\eta_\ell = 2-\sigma$ and $c$ is a constant. Accordingly, we can derive the leading scaling term of $M^2$, $\chi_k$ and $\xi$ in the LRO phase for $\sigma < 2$, which scales as $M^2 \sim M^2_0 + L^{-\eta_\ell}$, $\chi_k \sim L^{2-\eta_\ell}$ and $\xi \sim L^{1 + \eta_\ell/2}$. In the marginal case of $\sigma = 2$, however, the exact scaling form of the correlation function is not straightforward to derive. Nevertheless, it is natural to expect logarithmic corrections in this case as the anomalous dimension $\eta_{\ell}$ vanishes and the LR and SR terms become degenerate~\cite{wegner1972, wegner1973}. Hence, we conjecture that, at $\sigma = 2$, $M^2 \sim M_0^2 + \ln(L/L_0)^{\hat{\eta_\ell}}$, $\chi_k \sim L^{2}\ln(L/L_0')^{\hat{\eta}_\ell}$ and $\xi \sim L \ln(L/L_0'')^{-\hat{\eta}_\ell/2}$. Here, $\hat{\eta}_\ell$ is the exponent of the logarithmic correction, and $L_0$, $L_0'$, $L_0''$ are non-universal constants. See the Supplemental Material (SM) for detailed derivation and analysis~\cite{supp}.

\begin{table}[b]
    \centering
\begin{tabularx}{\linewidth}{X|XXX|X}
    \toprule
     $\sigma$ & $\beta_c$  & $1/\nu$ &      $\eta$     & $\eta_\ell$ \\
    \hline
      1.250   & 0.59961(2) & 0.95(2) &      0.743(10)  & 0.747(7) \\
      1.750   & 0.68380(3) & 0.66(4) &      0.329(14)  & 0.250(4) \\
      1.875   & 0.70737(4) & 0.60(2) &      0.288(10)  & 0.122(9) \\
      2.000   & 0.7315(2)  & 0.50(3) &      0.25(1)    & 0 ($1/\ln L$)        \\
    \bottomrule
\end{tabularx}
\caption{Critical point $\beta_c$ and critical exponents of LR XY model for various $\sigma$ in the non-classical regime. Here, $1/\nu$ is the correlation length exponent, and $\eta$ is the anomalous dimension. The decaying exponent $\eta_\ell$ of the low-$T$ correlation function is theoretically predicted to be $\eta_\ell=2-\sigma$. Note that, for $\sigma=2$, the estimates are seemingly consistent with $1/\nu=1/2$ and $\eta=1/4$, and $\eta_\ell = 0$ with a multiplicative logarithmic correction.}
\label{table_LRXY_fitting}
\end{table}

The upper-left insets of Fig.~\ref{fig_2_T_scan}, showing $(\xi/L)^2$ as a function of $\ln L$, demonstrate distinctive low-$T$ scaling behaviors of $\xi/L$ for different $\sigma$ values. For $\sigma = 2$, the data points can be well-described by straight lines of $\ln L$, which confirms the predicted scaling behavior and indicates $\hat{\eta}_\ell = -1$. On the other hand, for $\sigma = 1.75$, the bending-up curvatures mean that divergences of $\xi/L$ are faster than the logarithmic growth. The least-squares fit by $(\xi/L)^2 = c+ L^{\eta_\ell} (a+bL^{-1})$, with constants $a$, $b$ and $c$, gives $\eta_\ell =0.250(4)$, well consistent with the theoretical prediction. See SM for details of the fit~\cite{supp}, and values of $\eta_\ell$ are given in Table~\ref{table_LRXY_fitting}. By contrast, for $\sigma = 3$, $\xi/L$ quickly converges to a constant with increasing $L$. Direct evidence of LRO for $\sigma \le 2$ and $T < T_c$ is presented in the bottom-right insets of Fig.~\ref{fig_2_T_scan} by showing the low-$T$ scaling behavior of $M^2$. For $\sigma = 1.75$, extrapolating $M^{2}$ versus $L^{-\eta_\ell}$, with $\eta_\ell = 0.25$, illustrate finite magnetization in the $L\to\infty$ limit. For $\sigma = 2$, we fit the FSS ansatz $M^{2}(L) \sim c + a \ln(L/b)^{-1}$ and the extrapolation in the limit $L \rightarrow \infty$ demonstrate ferromagnetic phase. Analysis of the specific heat-like quantity, included in SM~\cite{supp}, also illustrates the second-order transitions for $\sigma \leq 2$. Our results provide compelling evidence that as long as $\sigma \leq 2$, the LR XY model enters a ferromagnetic phase through a second-order phase transition. This finding, however, is inconsistent with the phase diagram suggested in Ref.~\cite{giachetti2021, giachetti2022}, where the BKT transition is predicted to persist for $ 1.75 < \sigma \le 2$.

\begin{figure}[t]
    \centering
    \includegraphics{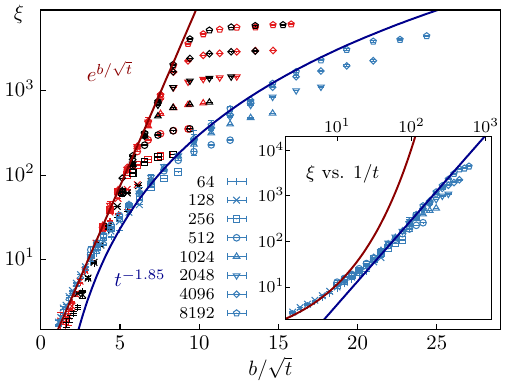}
    \vspace*{-6mm}
    \caption{Deviation of correlation length growth at $\sigma=2$ from the BKT scaling. The main panel shows a semi-logarithmic plot of $\xi$ as a function of $b/\sqrt{t}$ for various $L$ at $\sigma=2$ (blue dots), $3$ (red dots) and NN XY case (black dots), where the reduced temperature is $t=(T-T_c)/T_c$, and $b = 1, 1.25, 1.625$ respectively. For the $\sigma = 3$ and NN XY case, the linear behavior of $\xi$ demonstrates an exponential growth of $\xi$, characterizing the BKT transition. However, for $\sigma=2$, the growth of $\xi$ deviates more and more from the BKT behavior as the system approaches the critical point. The inset shows a double-log plot of $\xi$ versus $1/t$ for $\sigma = 2$, revealing a power-law behavior of $\xi$, thus highlighting the second-order phase transition.}
    \label{fig_3_RG}
\end{figure}

\begin{figure}[t]
    \centering
    \includegraphics[width=\columnwidth]{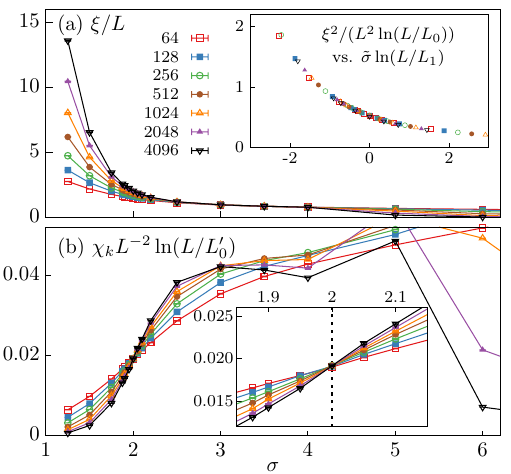}
    \vspace*{-6mm}
    \caption{Low-$T$ transitions at $\sigma = 2$ with $T=1$. (a) $\xi/L$ vs. $\sigma$ for various $L$. The system enters the LRO phase when $\sigma\le2$. The inset shows good data collapse of $\xi/(L \ln(L/L_0)^{1/2})$ vs. $\tilde{\sigma}\ln(L/L_1)$, where $\tilde{\sigma} = \sigma - 2$, $L_0 = 2.9$ and $L_1 = 3$. (b) $\chi_{k} L^{-2}\ln(L/L_0^{\prime})$ vs. $\sigma$ for various $L$, with $L_0^{\prime}=2.9$. The scaled $\chi_k$ curves have a clear crossing point at $\sigma = 2$ as demonstrated in both panel (b) and its inset.}
    \label{fig_3_sigma_scan}
\end{figure}

To resolve this inconsistency, revealing the type of phase transition for $\sigma=2$ becomes rather crucial. Note that our simulations are already up to $L=8192$; it is extremely difficult to improve the precision of critical exponents further. Hence, we adopt an alternative route by investigating the growth of correlation length $\xi$ as it approaches $T_c$. In the context of RG, near a BKT fixed point, $\xi$ exhibits an exponential divergence, $\xi \sim \exp{\left(b/\sqrt{t}\right)}$, where $t$ denotes the reduced temperature $t = (T_c - T)/T_c$ and $b$ is a non-universal constant~\cite{chen2022}. Conversely, $\xi$ diverges algebraically, $\xi \sim t^{-\nu}$, near a second-order transition. We first accurately determine the critical points $T_c(\sigma=2)=1.3671(4)$ and $T_c(\sigma=3)=1.109(2)$ by FSS analysis of $\xi/L$, and then study the growth of $\xi$ as a function of $t$. We plot $\xi$ against $b/\sqrt{t}$ on a semi-log scale for various $L$ for both $\sigma = 2$, $3$ and the NN case, as shown in Fig.~\ref{fig_3_RG}. For $\sigma=3$ and the NN case, data points of different $L$s converge onto a single linear trajectory consistent with the typical BKT divergence of $\xi$. For $\sigma=2$, when sufficiently away from $T_c$, the correlation length behaves seemingly like that for $\sigma=3$. However, as $t$ approaches $0$, the behavior of $\xi$ becomes increasingly different from SR cases and clearly distinct from the exponential growth, suggesting a different universality class. In contrast, the log-log plot in the inset shows $\xi$ can be well-described by an algebraic scaling $t^{-1.85}$. Note that $\nu=1.85$ slightly differs from the central value of the FSS fitting results $1/\nu = 0.50(3)$ but is still within two error bars. The $\xi$ growth for $\sigma =1.875$ also follows a power-law behavior as $t^{-1.62}$ with an exponent $\nu =1.62$ clearly different from that for $\sigma=2$ (see SM~\cite{supp}). These results strongly suggest that instead of being BKT-type, the phase transition at $\sigma = 2$ is a second-order transition, thus precluding the scenario proposed in Ref.~\cite{giachetti2021}.

The previous analysis demonstrates the logarithmic behavior of $\xi/L$ and $\chi_k$ in the low-$T$ phase at $\sigma = 2$. Hence, to further explore the low-$T$ physics at $\sigma=2$, we fix the temperature at $T=1.0$, which is below the critical point $T_c(\sigma=2)=1.3671(4)$ but sufficiently higher than the ground state, and study the behaviors of $\xi$ and $\chi_k$ as a function of $\sigma$. Fig.~\ref{fig_3_sigma_scan} (a) shows that the system undergoes three phases as $\sigma$ decreases. It first enters the QLRO phase from the disordered paramagnetic phase via a BKT transition at $\sigma \approx 4.0$; as $\sigma$ further declines, $\xi/L$ curves begin diverging near $\sigma = 2$, indicating the transition into LRO phases. We also plot $\chi_k L^{-2}\ln(L/L_0')$ as a function of $\sigma$ for various $L$ in Fig.~\ref{fig_3_sigma_scan} (b), with a constant $L_0'=2.9$. These curves exhibit an intersection at $\sigma = 2$, consistent with theoretical predictions.  A zoom-in plot in the inset better displays this crossing. Moreover, considering the logarithmic corrections at $\sigma = 2$ and $T < T_c$, we conjecture the scaling of $\xi$ near $\sigma = 2$ as $\xi = L \ln(L/L_0)^{\frac{1}{2}} g(\tilde{\sigma} \ln\left(L/L_1\right))$, where $\tilde{\sigma} = \sigma - 2$, $g$ is an universal scaling function and $L_0$, $L_1$ are unknown constants. As shown in the inset, the scaled $\xi$ data points collapse on the same curve, further supporting the $\sigma_* = 2$ scenario. See SM for more detail on the low-$T$ properties~\cite{supp}. Finally, in the thermodynamic limit, the magnetization density is finite in the LRO phase while vanishing in the QLRO phase, manifesting as a discontinuity in the order parameter at $\sigma = 2$. The QLRO phase does not exist at $T = 1$ when $\sigma \le 2$, inconsistent with the predicted low-$T$ behavior in Ref.~\cite{giachetti2021}.

\textit{Conclusion and Outlook}.---
Our results reveal that the 2D LR XY model enters a ferromagnetic phase at low temperatures through a second-order transition for $\sigma \leq 2$; in other words, the threshold value is $\sigma_* = 2$. The low-$T$ scaling behaviors are consistent with theoretical predictions. The power-law growth of $\xi$ near the critical point further demonstrates that the phase transition at $\sigma = 2$ is second-order, excluding the scenario predicted in Ref~\cite{giachetti2021}. Finally, for $\sigma = 2$ and $T < T_c$, the observed multiplicative logarithmic corrections also indicate the marginal nature of this point.

Preliminary investigations for the 2D LR Heisenberg model demonstrate that the algebraic interaction would induce a long-range ordered ferromagnet as long as $\sigma \leq 2$, while the system is in the disordered paramagnetic phase for $\sigma>2$ and $T>0$~\cite{WIP}. These combined messages make us conjecture that, for all the O($\mathcal{N}$) spin models, including the Ising model that has been extensively in the literature, the threshold value between the non-classical and the SR regime is always $\sigma_* =2$ and, thus, the Sak's criterion is most probably irrelevant. Undergoing studies are taken to test our conjecture. The success of this work suggests that, instead of simply improving over the estimate of critical exponents, one can study the system in an extended parameter space--e.g., the geometric structures of the Ising model and the self-avoid random walk (SAW), which corresponds to $\mathcal{N} \to 0$ limit of the O($\mathcal{N}$) spin model~\cite{fang2021}. 
In addition, the topology of our phase diagram differs from that of the LR quantum XXZ chain~\cite{giachetti2022, maghrebi2017}, which implies that the direct mapping~\cite{mattis1984} might be invalid here, posing an open question about the correspondence between LR classical and LR quantum model.
Finally, we emphasize that our work may be of timely application in cutting-edge experiments, like trapped ions and Rydberg-atom arrays, that are of LR interactions.

\begin{acknowledgments}
We thank Sheng Fang for initiating this project.
We acknowledge the support by the National Natural Science Foundation of China (NSFC) under Grant No. 12204173 and No. 12275263, and the Innovation Program for Quantum Science and Technology (under Grant No. 2021ZD0301900). YD is also supported by the Natural Science Foundation of Fujian Province 802 of China (Grant No. 2023J02032).
\end{acknowledgments}

\bibliography{long_range_XY}

\end{document}

\newcommand{\zj}[1]{\textcolor{blue}{ZJ: #1}}

%%%%%% reset %%%%%%
\clearpage
\onecolumngrid
\setcounter{equation}{0}
\setcounter{figure}{0}
\setcounter{table}{0}
\setcounter{page}{1}
\renewcommand{\thetable}{S\Roman{table}}
\renewcommand{\theequation}{S\arabic{equation}}
\renewcommand{\thefigure}{S\arabic{figure}}
%%%%%% reset %%%%%%

\begin{center}
{\large \bf Supplemental Material for\\[0.2em]
``Two-dimensional XY Ferromagnet Induced by Long-range Interaction''}\\[1em]
\end{center}

%%%%%%%%%%%%%%%%%%%%%%%%%%%%%%%%%%%%%%%%%%%%%%%%%%%%%%%%%%%%%%%%%%%%%%%%%%%
%%%%%%%%%%%%%%%%%%%%%%%%%%%%%%%%%%%%%%%%%%%%%%%%%%%%%%%%%%%%%%%%%%%%%%%%%%%
%%%%%%%%%%%%%%%%%%%%%%%%%%%%%%%%%%%%%%%%%%%%%%%%%%%%%%%%%%%%%%%%%%%%%%%%%%%
\section{The growth of the correlation length}
\begin{figure}[tb]
\centering
\includegraphics[width=\linewidth]{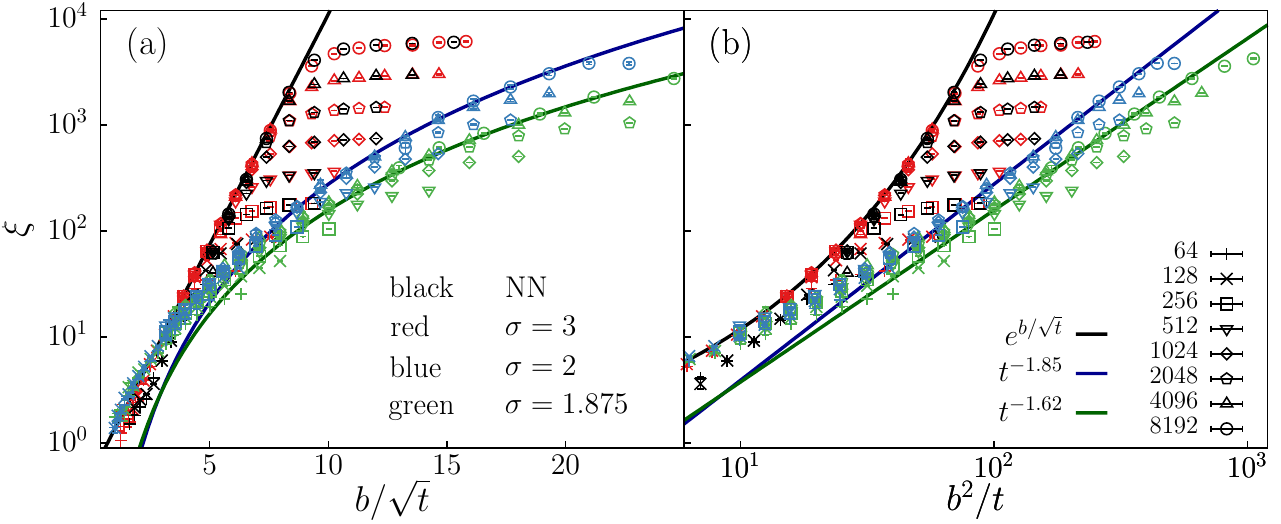}	
\caption{The growth of the correlation length $\xi$ with respect to the reduced temperature $t = (T - T_c)/T_c$ is shown for $\sigma = 1.875$ (green dots), $2$ (blue dots), $3$ (red dots), and the NN case (black dots). (a) The semi-logarithmic plot of $\xi$ versus $b/\sqrt{t}$, where $b=1$ for $\sigma=1.875$ and $\sigma=2$, and $b=1.25$ for $\sigma=3$ and $b=1.625$ for the NN case. The straight black line indicates exponential growth, suggesting that $\xi \sim \exp(b/\sqrt{t})$ for $\sigma = 3$ and the NN case. (b) The log-log plot of $\xi$ versus $b^2/t$. The straight dark-green and dark-blue lines indicate a power-law growth, with $\xi \sim t^{-\nu}$ for $\sigma = 1.875$ and $2$, respectively.}
\label{fig:rg_flow}
\end{figure}

In Fig.~\ref{fig:rg_flow}, we show the growth of the correlation length $\xi$ to demonstrate the universalities for different $\sigma$ values, identifying the threshold between short-range and long-range universalities as $\sigma_* = 2$.

In Fig.~\ref{fig:rg_flow}(a), a semi-log plot of $\xi$ versus $b/\sqrt{t}$ is presented. Here, $t = (T - T_c)/T_c$ and $b = 1.25, 1.625$ for $\sigma=3$ and the nearest-neighbor (NN) case, respectively, while $b = 1$ for others. The straight black line demonstrates that the correlation length follows $\xi \sim \exp(b/\sqrt{t})$ for $\sigma=3$ and the NN case, characterizing a typical BKT transition. Notably, the data collapse well, indicating that both $\sigma=3$ and NN cases belong to the universality class of the 2D short-range XY model. Moreover, two curved lines, colored dark green and dark blue, indicate a power-law behavior for $\sigma=2$ and 1.875, respectively. This behavior can be seen more clearly in Fig.~\ref{fig:rg_flow}(b). Additionally, as $b/\sqrt{t}$ increases, the system size truncates the correlation length, resulting in plateaus.

As for Fig.~\ref{fig:rg_flow}(b), we present a log-log plot of $\xi$ versus $b^2/t$, where the y-axis represents the square of $b/\sqrt{t}$, as in Fig.~\ref{fig:rg_flow}(a). The straight dark-green and dark-blue lines indicate a power-law growth, $\xi \sim t^{-\nu}$, for $\sigma=1.875$ and 2, implying a continuous phase transition. Moreover, the slope of these lines allows us to extract the critical exponent $\nu$. Notably, these values are consistent with those listed in our main text, supporting our assertion that $\sigma_* = 2$.

%%%%%%%%%%%%%%%%%%%%%%%%%%%%%%%%%%%%%%%%%%%%%%%%%%%%%%%%%%%%%%%%%%%%%%%%%%%
%%%%%%%%%%%%%%%%%%%%%%%%%%%%%%%%%%%%%%%%%%%%%%%%%%%%%%%%%%%%%%%%%%%%%%%%%%%
%%%%%%%%%%%%%%%%%%%%%%%%%%%%%%%%%%%%%%%%%%%%%%%%%%%%%%%%%%%%%%%%%%%%%%%%%%%
\section{Low temperature properties}
To explore the properties of the low-temperature phase, we conducted simulations for $\sigma=1.25, 1.75, 1.875, 2, 3$ at $\beta=1, 2, 4, 8$, where $\beta$ is the inverse temperature. The $\sigma=3$ case is included for comparison. The behaviors of $M^2$ and $\chi_k$ reveal the system's ferromagnetic nature and algebraically decaying correlation function at low temperatures, which we will elaborate on in detail later. Additionally, we present some fitting details of $\xi/L$, simplifying our presentation to fitting tables only for $\sigma=1.75$ and $2$ at $\beta=1$ and $4$.

%%%%%%%%%%%%%%%%%%%%%%%%%%%%%%%%%%%%%%%%%%%%%%%%%%%%%%%%%%%%%%%%%%%%%%%%%%%
\subsection{The existence of long-range order}

In this subsection, we demonstrate that for $\sigma \leqslant 2$, the system exhibits spontaneous magnetization, and continuous symmetry is broken. The fitting of $M^2$ can reveal the presence of spontaneous magnetization. If it does not exist, $M^2$ exhibits a power-law behavior as a function of $L$, allowing us to fit $M^2$ using the formula:
\begin{equation}\label{LE3}
    M^2=L^{-\eta_\ell}(a_0+b_1L^{-\omega})
\end{equation}
where $\eta_\ell$ is a special exponent in low temperature, which only depends on $\sigma$, and we reckon that $\eta_\ell=2-\sigma$, which will be explained in the next subsection. Otherwise, $M^2$ can be fitted to the formula:
\begin{equation}\label{LE4}
    M^2=g_0+L^{-\eta_\ell}(a_0+b_1L^{-\omega}),
\end{equation}
where $g_0$ represents the strength of spontaneous magnetization and the term $b_1L^{-\omega}$ is related to finite-size correction. Specifically, for $\sigma=2$ at low temperatures, as mentioned in the main text and the subsequent subsection, $\chi_k$ exhibits logarithmic behavior: $\chi_k\sim a_0L^2(\ln{L}+c_1)^{\hat\eta_\ell}$. According to the fitting results presented in the following subsection, we assume $\hat\eta_\ell=-1$. Thus, in the absence of spontaneous magnetization and in consistency with $\chi_k$, $M^2$ is expected to behave as:
\begin{equation}\label{LE5p}
    M^2=a_0(\ln{L}+c_1)^{\hat\eta_\ell},
\end{equation}
otherwise should behave as:
\begin{equation}\label{LE5}
    M^2=g_0+a_0(\ln{L}+c_1)^{\hat\eta_\ell}.
\end{equation}

\begin{figure}[t]
    \centering
    \includegraphics[scale=0.6]{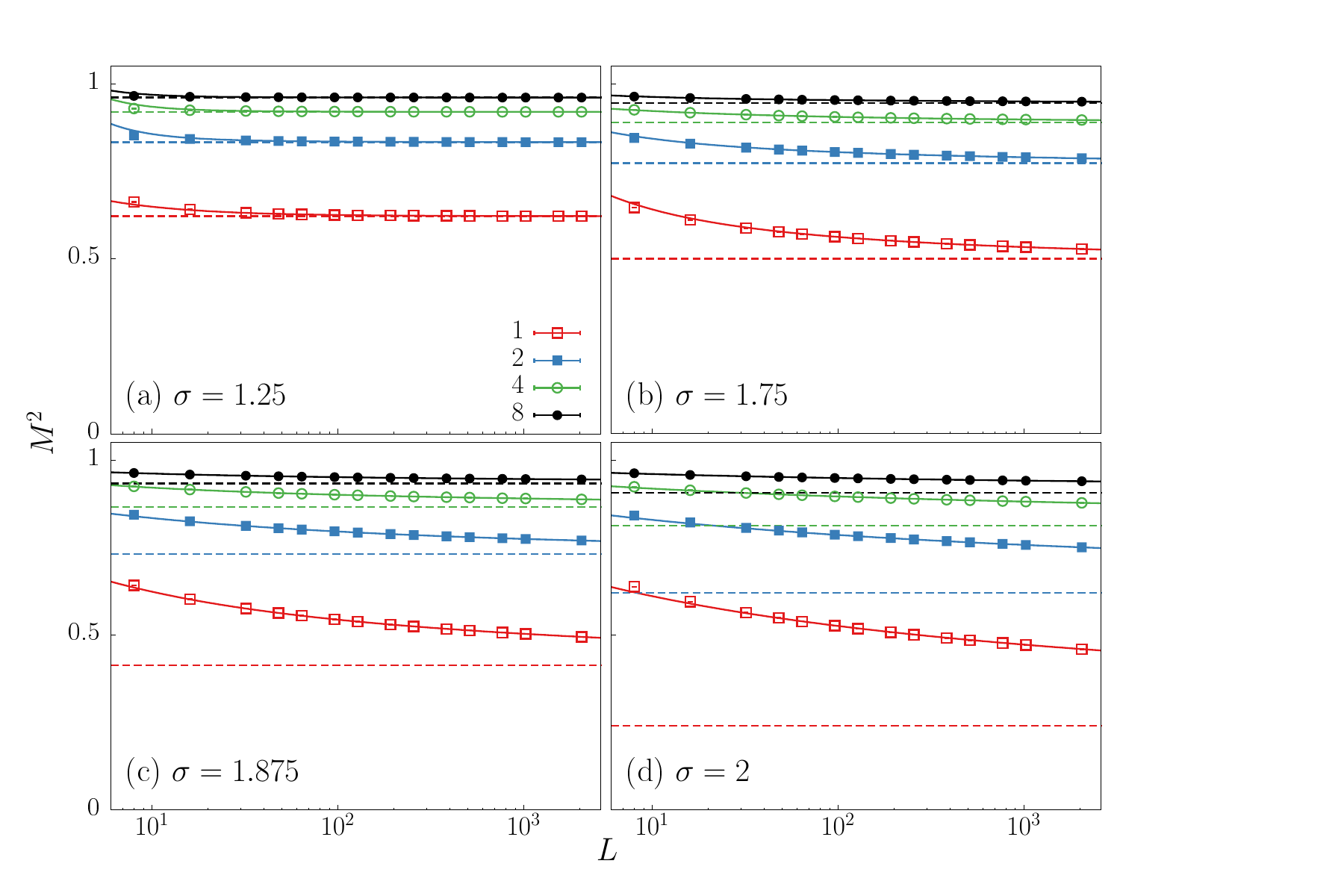}
    \caption{$M^2$ versus $L$ for $\sigma=1.25, 1.75, 1.875, 2$ at different temperatures, i.e., $\beta=1, 2, 4, 8$. The solid lines represent the curves fitted to the data for the corresponding temperatures, and the fitting formulas are given by Eq.\eqref{LE4} and Eq.\eqref{LE5}, respectively, for $\sigma<2$ and $\sigma=2$. The dashed lines represent the final value of $M^2$ in the limit of infinite system size, as determined by the fitting results.}
    \label{LTF3}
\end{figure}

\begin{figure}[htbp]
    \centering
    \includegraphics[scale=0.68]{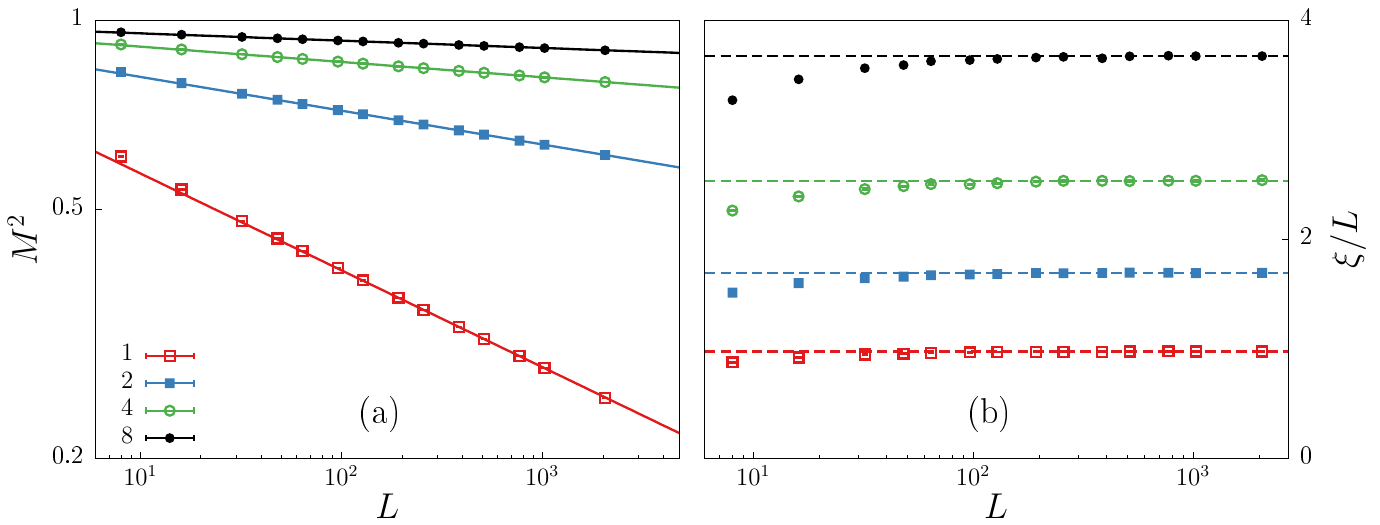}
    \caption{$M^2$ (a) and $\xi/L$ (b) versus $L$ for $\sigma=3$. Double logarithmic coordinates are adopted in (a), and only the horizontal axis is on a logarithmic scale in (b).}
    \label{LTF4}
\end{figure}

We first attempted to fit $M^2$ using Eq.\eqref{LE3} and Eq.\eqref{LE5p} for $\sigma<2$ and $\sigma=2$, respectively. However, we were unable to obtain appropriate fitting results. In other words, regardless of how we adjusted $\omega$ and $L_{\rm min}$ (the minimal $L$ used for fitting), $\chi^2/$DF remained large ($\gg 1$), where DF refers to the degree of freedom in the fitting. Subsequently, for $\sigma<2$ and $\sigma=2$, we proceeded to fit $M^2$ using Eq.\eqref{LE4} and Eq.\eqref{LE5}, with the results shown in Table~\ref{LT3} and Table~\ref{LT4}, respectively. Figure~\ref{LTF3}(a) clearly demonstrates the fitting results for $M^2$ for $\sigma=1.25, 1.75, 1.875, $ and $2$ at different temperatures, where high fitting quality suggests the existence of spontaneous magnetization at low temperatures. For comparison, we also fit $M^2$ to Eq.\eqref{LE3} for the case of $\sigma=3$, which yielded satisfactory results. The improved fitting outcomes suggest the absence of spontaneous magnetization. Figure~\ref{LTF4} clearly shows the power-law relationship between $M^2$ and $L$ for the case of $\sigma=3$, consistent with the absence of spontaneous magnetization.

\begin{table}[!ht]
    \centering
    \caption{Fits of $M^2$ to Eq.~\eqref{LE4} for $\sigma=1.75$}
    \begin{tabular}{p{2cm}p{2cm}p{2cm}p{2cm}p{2cm}p{2cm}p{2cm}p{1.2cm}}
    \hline\hline
        $\beta$ & $L_{\rm min}$ & $g_0$ & $a_0$ & $b_1$ & $\omega$&$\chi^2/$DF \\ \hline
        1.0 &128 & 0.50006(8) & 0.1848(5) & 0.37(1) &0.75& 3.2/5 \\ 
        &192 & 0.49997(9) & 0.1855(6) & 0.35(1)&0.75 & 2.0/4 \\ 
        &256 & 0.4999(1) & 0.1861(7) & 0.33(2) &0.75& 1.2/3 \\ 
         4.0 & 32 & 0.88938(2) & 0.0453(1) & 0.0356(5)&0.45 & 4.4/8 \\ 
        &48 & 0.88936(2) & 0.0453(1) & 0.0353(7)&0.45 & 4.0/7 \\ 
        &64 & 0.88938(2) & 0.0452(2) & 0.0359(9)&0.45 & 3.3/6 \\ 
        &96 & 0.88939(3) & 0.0452(2) & 0.036(1)&0.45 & 3.2/5 \\ 
        &128 & 0.88935(3) & 0.0455(2) & 0.034(1)&0.45 & 1.6/4 \\ 
        &256 & 0.88933(6) & 0.0456(5) & 0.033(3)&0.45 & 1.5/3 \\ 
        \hline\hline
    \end{tabular}\label{LT3}
\end{table}

\begin{table}[!ht]
    \centering
    \caption{Fits of $M^2$ to Eq.~\eqref{LE5} for $\sigma=2$}
    \begin{tabular}{p{2cm}p{2cm}p{2cm}p{2cm}p{2cm}p{1.2cm}}
    \hline\hline
        $\beta$ & $L_{\rm min}$ & $g_0$ & $a_0$ & $c_1$ & $\chi^2/$DF \\ \hline
        1.0&128 & 0.241(2) & 2.85(5) & 5.4(1) & 4.6/5 \\ 
        &192 & 0.239(3) & 2.90(7) & 5.5(1) & 3.7/4 \\ 
        &256 & 0.240(4) & 2.9(1) & 5.5(2) & 3.7/3 \\ 
        &384 & 0.232(8) & 3.1(2) & 5.9(4) & 2.2/2 \\
        4.0&64 & 0.8117(5) & 0.92(1) & 6.33(8) & 3.3/7 \\ 
        &96 & 0.8116(7) & 0.92(1) & 6.3(1) & 3.3/6 \\ 
        &128 & 0.812(1) & 0.91(2) & 6.3(1) & 3.1/5 \\ 
        &192 & 0.812(1) & 0.92(3) & 6.4(2) & 3.1/4 \\ 
        \hline\hline
    \end{tabular}\label{LT4}
\end{table}

%%%%%%%%%%%%%%%%%%%%%%%%%%%%%%%%%%%%%%%%%%%%%%%%%%%%%%%%%%%%%%%%%%%%%%%%%%%
\subsection{The Goldstone mode}

In this subsection, we demonstrate the algebraic correlation function in the low temperature, i.e., $g(x)=g_0+c\cdot x^{-\eta_\ell}$, induced by the Goldstone mode excitation under continuous symmetry breaking. It can be shown by theory and numerical results that $\eta_\ell=2-\sigma$ for $\sigma<2$.

\textit{Theoretical derivation}.
Hamiltonian in momentum-space is written as:
\begin{equation}
\begin{aligned}
    \beta H=&\int{\frac{\mathrm d^dq}{(2\pi)^d}(\frac t2+\frac {K_2}{2} q^2+K_\sigma q^\sigma)\boldsymbol{\Psi}(\boldsymbol{q})\cdot\boldsymbol{\Psi}(-\boldsymbol{q})}+\\
    &\int\frac{\mathrm{d}^dq_1}{(2\pi)^d}\int\frac{\mathrm{d}^dq_2}{(2\pi)^d}\int\frac{\mathrm{d}^dq_3}{(2\pi)^d}u\boldsymbol{\Psi}(\boldsymbol{q_1})\cdot\boldsymbol{\Psi}(\boldsymbol{q_2})\cdot\boldsymbol{\Psi}(\boldsymbol{q_3})\cdot\boldsymbol{\Psi}(-\boldsymbol{q_1}-\boldsymbol{q_2}-\boldsymbol{q_3}).
\end{aligned}
\label{Hamiltonian}
\end{equation}
For $\sigma<2$, $K_\sigma q^\sigma$ is the leading term, and hence $\frac {K_2}{2} q^2$ can be ignored. Also, for the simplicity of computation, the Hamiltonian is transformed to:
\begin{equation}
    \beta H=\int{\frac{\mathrm d^dq}{(2\pi)^d}K_\sigma q^\sigma\boldsymbol{\Psi}(\boldsymbol{q})\cdot\boldsymbol{\Psi}(-\boldsymbol{q})}+\int\mathrm{d}^dx(\frac t2\Psi^2+u\Psi^4). 
    \label{Hamiltonian2}
\end{equation}

In the last subsection, we have shown that there exist spontaneous magnetization and continuous symmetry breaking in low temperatures for $\sigma\leqslant2$. Thus in the following, we adopt saddle point approximation and consider small fluctuation around it. Under saddle point approximation, the Hamiltonian becomes:
\begin{equation}
    \beta H=V(\frac t2{\overline{\Psi}}^2+u{\overline{\Psi}}^4),
\end{equation}
where $V$ refers to the system volume. Thus,
\begin{equation}
\overline{\Psi}=
\begin{cases}
    0 &, t>0\\
    \sqrt{\frac{-t}{4u}} &, t<0
\end{cases}
.
\label{Saddle}
\end{equation}
Since only the transverse fluctuation matters, we ignore the longitudinal fluctuation and treat the transverse fluctuation as a small amount. The spin field is written as (for simplicity, only consider the case of $d=2$):
\begin{equation}
    \boldsymbol{\Psi}(\boldsymbol{x})=\Psi_L(\boldsymbol{x})\boldsymbol{\hat{e}_l}+\Psi_T(\boldsymbol{x})\boldsymbol{\hat{e}_t}=\overline{\Psi}\boldsymbol{\hat{e}_l}+\Psi_T(\boldsymbol{x})\boldsymbol{\hat{e}_t}.
    \label{spinField}
\end{equation}
Its Fourier transformation is:
\begin{equation}
    \boldsymbol{\Psi}(\boldsymbol{q})=\int\mathrm d^dx~ \boldsymbol{\Psi}(\boldsymbol{x})e^{-i\boldsymbol{q}\cdot\boldsymbol{x}}={(2\pi)}^d\delta(\boldsymbol{q})\cdot\overline{\Psi}\cdot\boldsymbol{\hat e_l} + \Psi_T(\boldsymbol{q})\cdot \boldsymbol{\hat e_t}.
    \label{Fourier}
\end{equation}
By substituting Eq. \eqref{spinField}, \eqref{Saddle}, \eqref{Fourier} into Eq. \eqref{Hamiltonian2} (Note that all derivations are conducted under the condition of low temperature, hence $\overline{m}=\sqrt{\frac{-t}{4u}}$) and simplifying (Neglect small terms of third order and higher), we can obtain:
\begin{equation}
    \beta H=V(\frac t2{\overline{\Psi}}^2+u{\overline{\Psi}}^4)+\frac1V\sum_{\boldsymbol{q}}K_\sigma q^\sigma\cdot|\Psi_T(\boldsymbol{q})|^2.
\end{equation}
Thus the probability of a particular fluctuation configuration is given
by:
\begin{equation}
    P(\{\Psi_T(\boldsymbol{q})\})\propto e^{-\beta H}\propto \prod_{\boldsymbol{q}}\exp{(\frac {K_\sigma}{V}q^\sigma\cdot|\Psi_T(\boldsymbol{q})|^2)}.
\end{equation}
Hence, the two-point correlation function in the momentum space is:
\begin{equation}
    \langle\Psi_T(\boldsymbol{q})\Psi_T(\boldsymbol{q'})\rangle=\frac{\delta_{\boldsymbol{q},-\boldsymbol{q'}}V}{2K_\sigma q^\sigma},
\end{equation}
and in the real space is:
\begin{equation}
\begin{aligned}
    \langle\Psi_T(\boldsymbol{x})\Psi_T(\boldsymbol{x'})\rangle&=\frac1{V^2}\sum_{\boldsymbol{q}, \boldsymbol{q'}}\langle\phi(\boldsymbol{q})\phi(\boldsymbol{q'})\rangle e^{i\boldsymbol{q}\cdot\boldsymbol{x}+i\boldsymbol{q'}\cdot\boldsymbol{x'}}\\
    &=\frac1V\sum_{\boldsymbol{q}}\frac{e^{i\boldsymbol{q}(\boldsymbol{x}-\boldsymbol{x'})}}{2K_\sigma q^\sigma}\\
    &=\frac1{2K_{\sigma}}\int\frac{\mathrm{d}^dq}{(2\pi)^d}\frac{e^{i\boldsymbol{q}(\boldsymbol{x}-\boldsymbol{x'})}}{q^\sigma}.
\end{aligned}
\end{equation}
Consider the integration (Note that our derivation is under the condition of $\sigma<2=d$):
\begin{equation}
\begin{aligned}
    \int\frac{\mathrm{d}^dq}{(2\pi)^d}\frac{e^{i\boldsymbol{q}\cdot\boldsymbol{x}}}{q^\sigma}&=\int\frac{\mathrm d\Omega}{(2\pi)^d}\int\mathrm dq\frac{e^{iqx\cos\theta}}{q^{\sigma-d+1}}\\
    &=x^{\sigma-d}\int\frac{\mathrm d\Omega}{(2\pi)^d}\int\mathrm dy\frac{e^{iy\cos\theta}}{y^{\sigma-d+1}}\\
    &\sim x^{\sigma-d}.
\end{aligned}
\end{equation}
The correlation function:
\begin{equation}
\begin{aligned}
    g(x)&=\langle\boldsymbol{\Psi}(\boldsymbol{x})\cdot\boldsymbol{\Psi}(0)\rangle\\
    &=\overline{\Psi}^2+\langle\Psi_T(\boldsymbol{x})\cdot\Psi_T(0)\rangle\\
    &=\overline{\Psi}^2+c\cdot x^{\sigma-d}.
\end{aligned}
\end{equation}
Hence, for $\sigma<2$ at low temperatures, the correlation function takes on an algebraic form: $g(x)=g_0+c\cdot x^{-\eta_\ell}$, with $\eta_\ell=2-\sigma$. It's noteworthy that the derivation above is applicable only at low temperatures. Near or at the critical temperature, the system exhibits little to no spontaneous magnetization, and fluctuations play a significant role that cannot be considered negligible anymore. Therefore, the correlation function may exhibit different behaviors at low temperatures compared to the critical temperature, as will be seen in the subsequent analysis of $\chi_k$.

\textit{Fitting of $\chi_k$}.
\begin{figure}[t]
    \centering
    \includegraphics[scale=0.55]{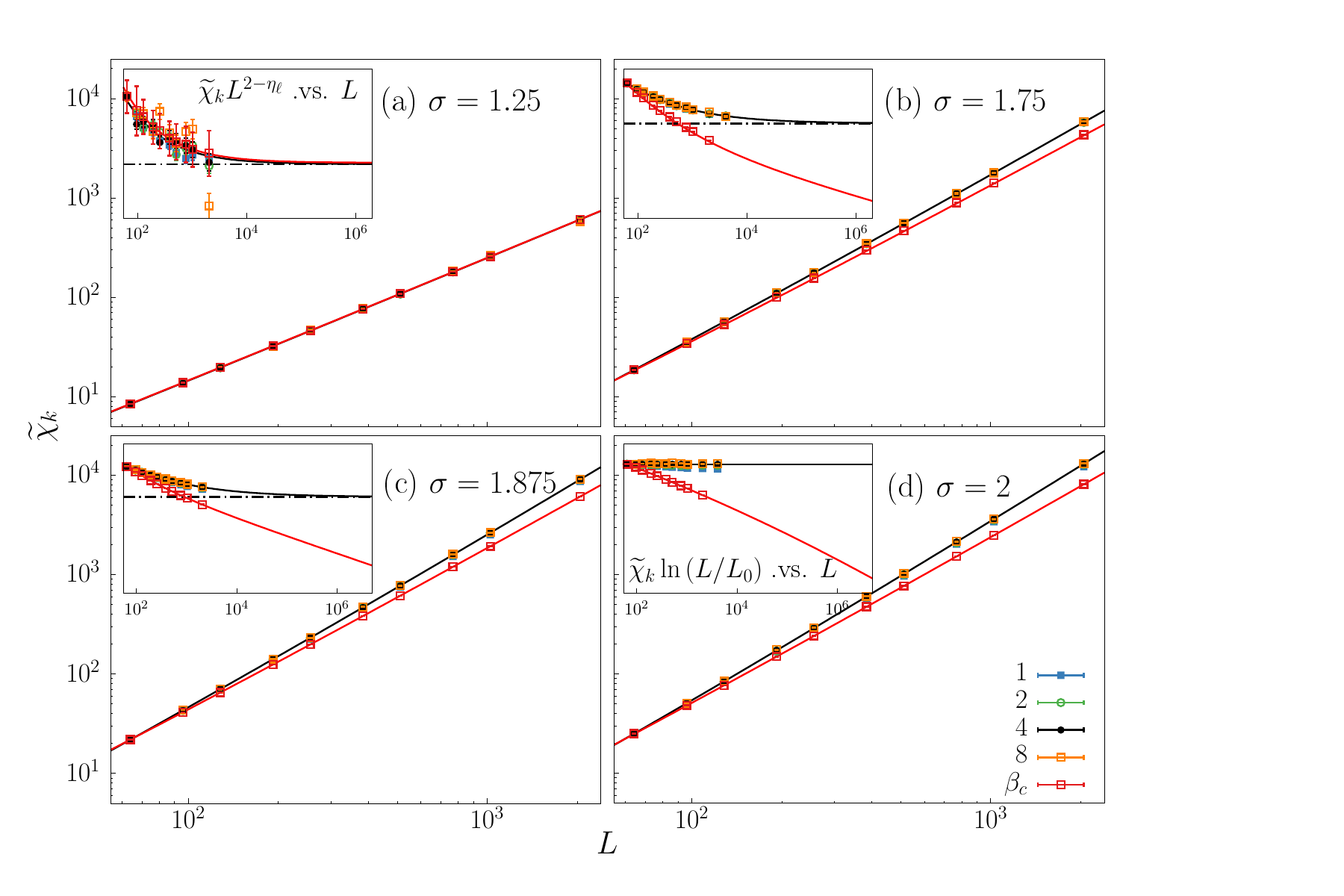}
    \caption{$\widetilde\chi_k$ versus $L$ at different temperatures for $\sigma$ = 1.25, 1.75, 2, and 3. $\widetilde\chi_k$ represents $\chi_k$ multiplied by a constant to make the first data point at different temperatures ($L=64$) overlap. Black and red lines represent the fitting curves of data at $\beta=4$ and $\beta=\beta_c$, respectively. The insets plot the relationship between $\widetilde\chi_kL^{2-\eta_\ell}$ and $L$ (in (a), (b), (c)) and the relationship between $\widetilde\chi_k\ln{(L/L_0)}$ and $L$ (in (d)), respectively. At low temperatures, the curves flatten out and finally converge to a constant, indicating the scaling behavior: $\chi_k\sim L^{2-\eta_\ell}$ for $\sigma<2$ and $\chi_k\sim\ln{(L/L_0)}$ for $\sigma=2$. However, for $\sigma=1.75, 1.875,$ and $2$, $\chi_k$ exhibits clearly different scaling behaviors at the critical point and low temperatures.}
    \label{LTF1}
\end{figure}
Considering the correlation function has the form: $g(x)=g_0+c\cdot x^{-\eta_\ell}$, then it can be shown that $\chi_k\sim L^{d-\eta_\ell}$:
\begin{equation*}
        \begin{aligned}
        \chi_k&=L^d|\boldsymbol{m_k}|^2\\
        &=\frac{L^d}{L^{2d}}(\sum_j\boldsymbol{S_j}\cdot e^{i\boldsymbol{k}\cdot\boldsymbol{x_j}})(\sum_i\boldsymbol{S_i}\cdot e^{-i\boldsymbol{k}\cdot\boldsymbol{x_i}})\\
        &=\frac1{L^{d}}\sum_{ij}(\boldsymbol{S_i\cdot S_j})\cdot e^{i\boldsymbol{k}\cdot(\boldsymbol{x_j}-\boldsymbol{x_i})}\\
        &=\frac1{L^d}\sum_i\sum_{\boldsymbol{x}}g(x)e^{i\boldsymbol{k\cdot x}}=\sum_{\boldsymbol{x}}g(x)e^{i\boldsymbol{k\cdot x}}=\sum_{\boldsymbol{x}}x^{-\eta_\ell}e^{i\boldsymbol{k\cdot x}} \propto L^{d-\eta_\ell}.
    \end{aligned}
\end{equation*}
Therefore, through $\chi_k$, we can study the exponent $\eta_\ell$. Figure~\ref{LTF1} plots $\widetilde\chi_k$ versus $L$, where $\widetilde\chi_k$ represents $\chi_k$ multiplied by a constant to make the first data point ($L=64$) at different temperatures overlap. 

As shown in Fig.~\ref{LTF1}, the relationship between $\chi_k$ and $L$ gradually approaches a power-law behavior, which indicates an algebraic decay of correlation function: $g(x)=g_0+c \cdot x^{-\eta_\ell}$. Moreover, $\widetilde\chi_k$ at different low temperatures almost remain overlapped, and small deviations may come from the finite size effect, suggesting that $\eta_\ell$ only depends on $\sigma$. For $\sigma=1.75,1.875$ and $2$, $\widetilde\chi_k$ at 
the critical point clearly separates with that at low temperatures, which indicates that the power-law decaying term of the correlation function has different exponents for low temperatures and the critical temperature. These numerical results are consistent with our theoretical derivation. To further support our theory, for $\sigma<2$, we fit the data to the formula:
\begin{equation}\label{LE1}
    \chi_k=L^{2-\eta_\ell}(a_0+b_1L^{-\omega})+c,
\end{equation}
where the constant term comes from the analytic part of the free energy and $\eta_\ell$ is fixed at $2-\sigma$. During the fitting, it's observed that $c$ always stays very close to 0. Thus we also try to fix $c=0$ in the fitting. Some results are presented in Table~\ref{LT1}. Actually for $\sigma=1.25, 1.75, 1.875$, data at all temperatures ($\beta=1,2,4,8$) can be fitted well to Eq.~\eqref{LE1}. 

\begin{figure}[t]
    \centering
    \includegraphics[scale=1]{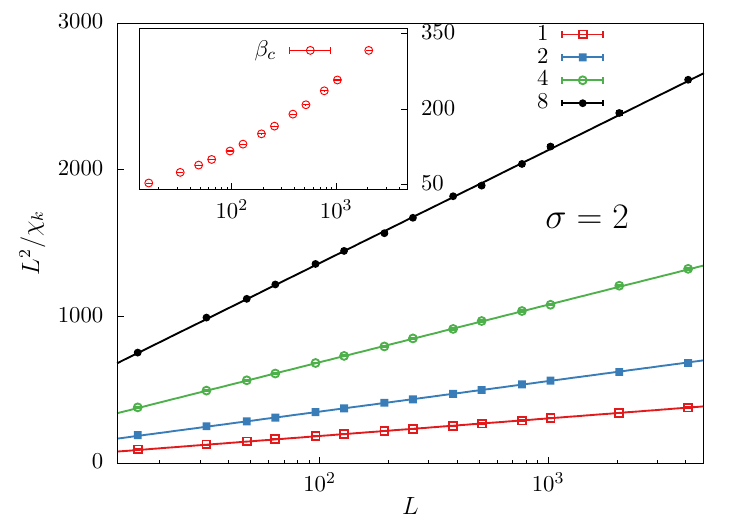}
    \caption{$L^2/\chi_k$ versus $L$ for $\sigma=2$ at $\beta=1, 2, 4, 8$. The horizontal axis is on a logarithmic scale. The solid lines represent straight lines fitted to the data at corresponding temperatures. The good fit demonstrates that $L^2/\chi_k$ has a linear relationship with $\ln{L}$. The inset plots data points at the critical temperature, and its exponential growth behavior indicates a power-law relationship between $L^2/\chi_k$ and $L$.}
    \label{LTF2}
\end{figure}

For $\sigma=2$, in Eq.~\eqref{Hamiltonian}, two terms, i.e., $\frac {K_2}2q^2$ and $K_\sigma q^\sigma$, become degenerate which may cause logarithmic corrections. So, we first attempt to use $\chi_k=L^2(\ln{(L/L_0)})^{\hat\eta_\ell}(a_0+b_1L^{-\omega})$ for fitting. Then we find the term $b_1L^{-\omega}$ is redundant. After removing that term, we fit the data to the formula: 
\begin{equation}\label{LE2}
    \chi_k=a_0L^2(\ln{L}+c_1)^{\hat\eta_\ell}+c, 
\end{equation}
where $c$ is also found to be very close to 0, it is fixed at 0 for the fitting. Some results are displayed in Table~\ref{LT2}. It is apparent that at different temperatures (including $\beta=2, 8$, which are not shown here), the value of $\hat\eta_\ell$ obtained from the fitting consistently hovers around -1. For simplicity, we assume $\hat\eta_\ell$ precisely equals -1. Based on this assumption, we re-fit the data with $\hat\eta_\ell$ fixed at -1, and the results are also presented in Table~\ref{LT2}. As indicated by the table, the satisfactory fitting results support our hypothesis. Figure~\ref{LTF2} visually demonstrates that at various low temperatures, $L^2/\chi_k$ and $\ln{L}$ have a linear relationship, specifically, $L^2/\chi_k\sim a\ln{L}+b$. Consequently, $\chi_k\sim a_0L^2(\ln{L}+c_1)^{-1}$. However, at the critical point, in contrast to the behavior observed at low temperatures, $L^2/\chi_k$ and $L$ demonstrate a power-law relationship.

\begin{table}[!ht]
    \centering
    \caption{Fits of $\chi_k$ to Eq.~\eqref{LE1} for $\sigma=1.75$}
    \begin{tabular}{p{2cm}p{2cm}p{2cm}p{2cm}p{2cm}p{2cm}p{2cm}p{1.2cm}}
    \hline\hline
    $\beta$ & $L_{\rm{min}}$ & $a_0$ & $b_1$ & $\omega$ & $c$ & $\chi^2$/DF \\ \hline
    1.0 & 32 & 0.0100(1) & 0.060(4) & -0.69(2) & -0.00016(2) & 7.5/9 \\ 
     & 48 & 0.0097(2) & 0.052(5) & -0.64(3) & -0.00013(2) & 5.2/8 \\ 
     & 64 & 0.0094(2) & 0.043(4) & -0.58(4) & -0.00010(3) & 3.2/7 \\ 
     & 96 & 0.0096(4) & 0.05(1) & -0.62(6) & -0.00012(3) & 3.0/6 \\ 
     & 128 & 0.0095(5) & 0.05(1) & -0.6(1) & -0.00012(5) & 3.0/5 \\ 
     & 192 & 0.009(1) & 0.03(1) & -0.5(2) & -0.0001(1) & 2.6/4 \\ 
     & 64 & 0.00829(6) & 0.032(1) & -0.47(1) & 0 & 7.2/8 \\ 
     & 96 & 0.00823(8) & 0.030(2) & -0.45(2) & 0 & 6.0/7 \\ 
     & 128 & 0.0082(1) & 0.027(2) & -0.43(2) & 0 & 4.7/6 \\ 
     & 192 & 0.0080(1) & 0.023(2) & -0.39(3) & 0 & 2.8/5 \\ 
     & 256 & 0.0081(1) & 0.026(4) & -0.42(4) & 0 & 2.4/4 \\ 
    4.0 & 32 & 0.00252(7) & 0.0097(7) & -0.57(3) & -0.000036(8) & 3.6/9 \\ 
     & 48 & 0.0025(1) & 0.009(1) & -0.55(5) & -0.00003(1) & 3.5/8 \\ 
     & 64 & 0.0025(1) & 0.010(2) & -0.59(7) & -0.00004(1) & 3.2/7 \\ 
     & 96 & 0.0023(2) & 0.007(1) & -0.43(9) & -0.00001(2) & 2.0/6 \\ 
     & 128 & 0.0024(2) & 0.008(3) & -0.5(1) & -0.00003(2) & 1.8/5 \\ 
     & 192 & 0.0020(9) & 0.005(2) & -0.3(2) & 0.00000(5) & 1.5/4 \\ 
     & 48 & 0.00213(2) & 0.0071(3) & -0.43(1) & 0 & 6.3/9 \\ 
     & 64 & 0.00212(3) & 0.0068(4) & -0.42(2) & 0 & 5.9/8 \\ 
     & 96 & 0.00206(2) & 0.0058(3) & -0.37(1) & 0 & 2.1/7 \\ 
     & 128 & 0.00206(3) & 0.0058(5) & -0.37(2) & 0 & 2.1/6 \\ 
     & 192 & 0.00202(5) & 0.0051(6) & -0.34(3) & 0 & 1.5/5 \\ 
    \hline\hline
    \end{tabular}\label{LT1}
\end{table}

\begin{table}[!ht]
    \centering
    \caption{Fits of $\chi_k$ to Eq.~\eqref{LE2} for $\sigma=2$ }
    \begin{tabular}{p{2cm}p{2cm}p{2cm}p{2cm}p{2cm}p{1.2cm}}
    \hline\hline
        $\beta$ & $L_{\rm min}$ & $\hat\eta_\ell$ & $c_1$ & $a_0$ & $\chi^2/$DF \\ \hline 
        1.0 &48 & -0.99(2) & -0.8(1) & 0.0108(7) & 3.5/8 \\ 
        &64 & -0.98(3) & -0.8(1) & 0.011(1) & 3.5/7 \\ 
        &96 & -0.99(5) & -0.8(2) & 0.011(1) & 3.5/6 \\ 
        &128 & -0.94(6) & -1.0(3) & 0.009(1) & 2.8/5 \\ 
        &192 & -0.9(1) & -1.0(6) & 0.009(3) & 2.8/4 \\  
        & 96 & -1 & -1.06(1) & 0.01918(7) & 4.0/7 \\ 
        &128 & -1 & -1.06(2) & 0.01919(9) & 4.0/6 \\ 
        &192 & -1 & -1.09(2) & 0.01907(9) & 2.0/5 \\ 
        &256 & -1 & -1.10(3) & 0.0190(1) & 1.9/4 \\ 
        &384 & -1 & -1.13(5) & 0.0189(1) & 1.5/3 \\ 
        4.0 & 32 & -1.03(2) & -0.5(1) & 0.0063(4) & 4.9/9 \\ 
        &48 & -1.05(4) & -0.3(2) & 0.0067(7) & 4.6/8 \\ 
        &64 & -1.04(6) & -0.4(3) & 0.007(1) & 4.6/7 \\ 
        &96 & -1.13(9) & 0.1(5) & 0.008(2) & 3.5/6 \\ 
        &128 & -1.1(1) & 0.1(7) & 0.008(3) & 3.5/5 \\ 
        &192 & -1.1(2) & -0(1) & 0.007(4) & 3.4/4 \\ 
        &32 & -1 & -0.57(1) & 0.00587(2) & 5.4/10 \\ 
        &48 & -1 & -0.57(2) & 0.00587(2) & 5.4/9 \\ 
        &64 & -1 & -0.58(2) & 0.00585(3) & 4.9/8 \\ 
        &96 & -1 & -0.57(3) & 0.00586(4) & 4.8/7 \\ 
        &128 & -1 & -0.60(4) & 0.00583(5) & 4.3/6 \\ 
        &192 & -1 & -0.64(6) & 0.00579(6) & 3.5/5 \\ 
        &256 & -1 & -0.61(8) & 0.00582(8) & 3.2/4 \\ 
        \hline\hline
    \end{tabular}
    \label{LT2}
\end{table}

%%%%%%%%%%%%%%%%%%%%%%%%%%%%%%%%%%%%%%%%%%%%%%%%%%%%%%%%%%%%%%%%%%%%%%%%%%%
\subsection{Fitting of $\xi/L$}
The quantity $\xi/L$ can further reveal the properties of low-temperature phase. In this subsection, we provide a numerical estimate of $\eta_\ell$ through the fitting of $(\xi/L)^2$, and further demonstrate system's ferromagnetic nature at low temperature for the case of $\sigma\leqslant 2$.

\begin{figure}
    \centering
    \includegraphics[scale=0.52]{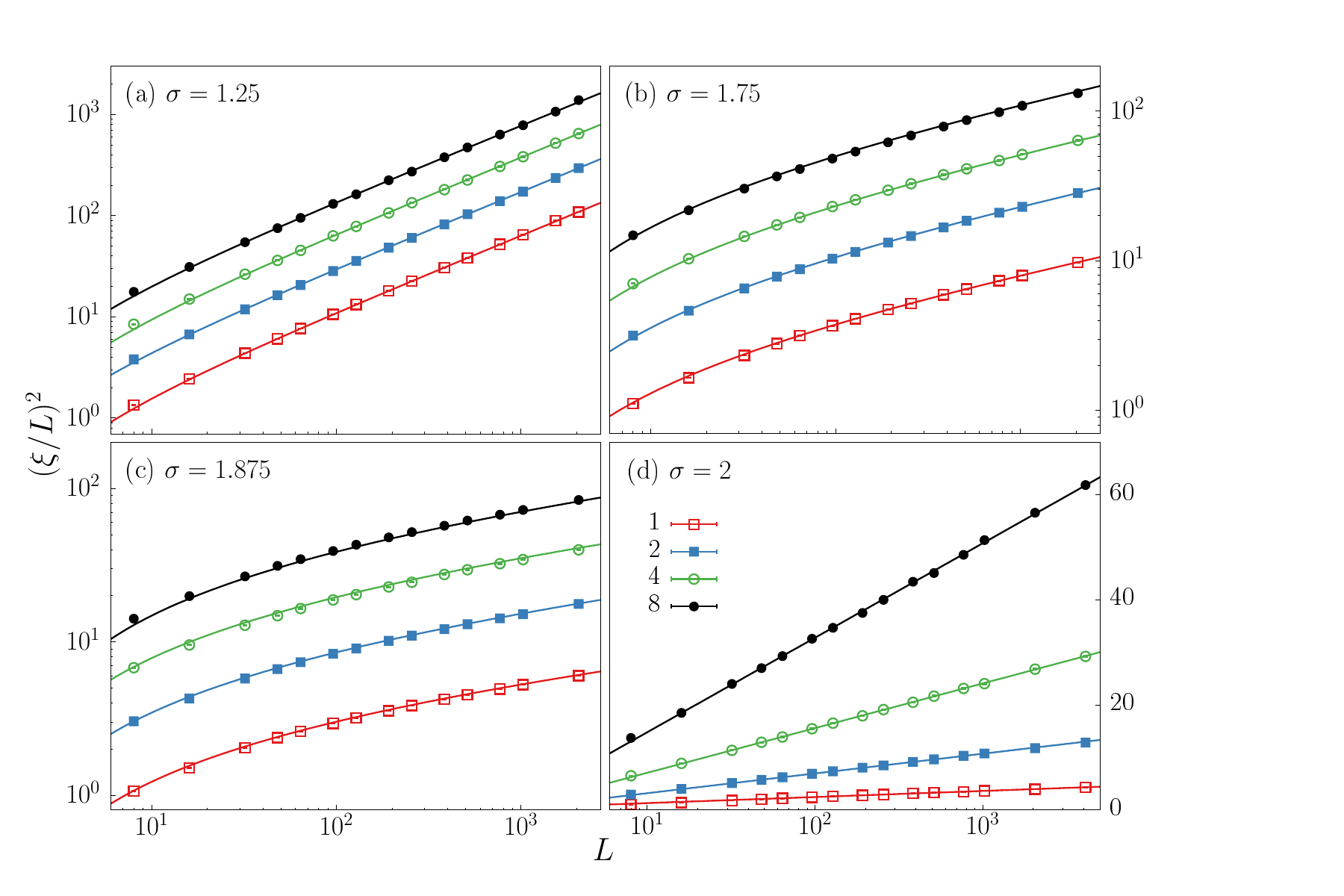}
    \caption{$(\xi/L)^2$ versus $L$ for $\sigma=1.25, 1.75, 1.875, 2$ at $\beta=1, 2, 4, 8$. Double logarithmic coordinates are used in (a), (b), and (c), and only the horizontal axis is on a logarithmic scale in (d). The solid lines represent fitting curves. The fitting formula is given by Eq. \eqref{LE6.5} for $\sigma<2$ and Eq. \eqref{LE7} for $\sigma=2$. From the graph, it is observed that as $L$ increases, $(\xi/L)^2$ tends to diverge, following a power-law behavior for $\sigma<2$ and a logarithmic behavior for $\sigma=2$. Error bars are within symbols if not visible.}
    \label{LTF5}
\end{figure}

As the system size increases, $\xi/L$ quickly reduces to 0 for the disordered phase, converges to a constant for the quasi-long-range-order phase, and diverges for the ferromagnetic phase. The insets of Figure 2 in the main text display the variation of $(\xi/L)^2$ with respect to $L$. As mentioned in the main text, $\xi/L = 1 /\left[2L \sin(|\bm{k}|/2)\right] \sqrt{\langle M^2 \rangle/\langle M_k^2 \rangle - 1}$. For $\sigma<2$, where $M_k^2\sim L^{-\eta_\ell}$ and $M^2\sim \text{Const}$, then $\xi/L\sim L^{\eta_\ell/2}$. Therefore, we fit $(\xi/L)^2$ to the formula:
\begin{equation}\label{LE6}
    (\xi/L)^2=L^{\eta_\ell}(a_0+b_1L^{-1})+c,
\end{equation}
where $c$ comes from the analytic part of freedom energy, and $b_1L^{-1}$ refers to the finite-size-correction term. Then we find $b_1L^{-1}$ is redundant, that is, removing the term merely changes the fitting results. Thus we ignore the correction term $b_1L^{-\omega}$ and fit $(\xi/L)^2$ to the formula:
\begin{equation}\label{LE6.5}
    (\xi/L)^2=c+aL^{\eta_\ell}
\end{equation}
at $\beta=1, 2, 4, 8$ for the cases of $\sigma=1.25, 1.75, 1.875$. Here, $c$ can be seen as a correction term with its exponent being $-\eta_\ell$. Table~\ref{LT5} presents the estimates of $\eta_\ell$ for different parameters. While for $\sigma=2$, $M^2_k\sim \left[\ln{(L/L_0)}\right]^{-1}$, $M^2\sim Const$, then $\xi/L\sim \left[\ln{(L/L_0)}\right]^{1/2}$. Hence $(\xi/L)^2$ is fitted to:
\begin{equation}\label{LE7}
    (\xi/L)^2=a_0(\ln{L}+c).
\end{equation}
Fitting results are shown in Table~\ref{LT6}. Figures \ref{LTF5} and \ref{LTF4}(b) illustrate the variation of $(\xi/L)^2$ with respect to $L$, as well as the fitting curves for $\sigma=$ 1.25, 1.75, 1.875, 2, and 3. The different behaviors of $\xi/L$ for $\sigma \leq 2$ and $\sigma > 2$ indicate distinct low-temperature characteristics. The diverging behavior of $\xi/L$ for $\sigma \leq 2$ further demonstrates the ferromagnetic nature at low temperatures.

\begin{table}[!ht]
    \centering
    \begin{threeparttable}
    \caption{Estimates of $\eta_\ell$ for different parameters are provided. The fitting formula used is Eq. \eqref{LE6.5}. The data in the 'low-T' column combines the data from the previous four temperatures.}
    \begin{tabular}{p{2cm}p{2cm}p{2cm}p{2cm}p{2cm}p{2cm}p{2cm}p{2cm}}
    \hline\hline
        $\sigma$&$\beta=1$ & $\beta=2$ & $\beta=4$ & $\beta=8$ & low-T & theory \\ \hline
        1.25&0.751(3)&0.754(4)& 0.751(5) & 0.743(7) & 0.747(7)& 0.75\\
        1.75 & 0.237(6) & 0.247(3) & 0.253(3) & 0.250(4) & 0.250(4) & 0.25\\
        1.875 & 0.098(8) & 0.121(8) & 0.123(7) & 0.121(9) & 0.122(9) & 0.125\\
        \hline\hline
    \end{tabular}\label{LT5}
    \end{threeparttable}
\end{table}

\begin{table}[!ht]
    \centering
    \caption{Fits of $(\xi/L)^2$ to \eqref{LE7} for $\sigma=2$}
    \begin{tabular}{p{2cm}p{2cm}p{2cm}p{2cm}p{1.2cm}}
    \hline\hline
        $\beta$ & $L_{\rm min}$ & $c$ & $a_0$ & $\chi^2/$DF \\ \hline
        1.0&128 & 0.29(5) & 0.503(4) & 6.3/6 \\ 
        &192 & 0.33(6) & 0.500(4) & 5.2/5 \\ 
        &256 & 0.41(7) & 0.494(5) & 3.2/4 \\ 
        4.0&64 & -0.38(2) & 1.922(4) & 4.3/9 \\ 
        &96 & -0.37(2) & 1.918(4) & 3.6/8 \\ 
        &128 & -0.38(3) & 1.920(6) & 3.5/7 \\ 
        &192 & -0.39(4) & 1.923(7) & 3.3/6 \\ 
        &256 & -0.35(5) & 1.916(9) & 2.5/5 \\ 
        \hline\hline
    \end{tabular}\label{LT6}
\end{table}
%%%%%%%%%%%%%%%%%%%%%%%%%%%%%%%%%%%%%%%%%%%%%%%%%%%%%%%%%%%%%%%%%%%%%%%%%%%
%%%%%%%%%%%%%%%%%%%%%%%%%%%%%%%%%%%%%%%%%%%%%%%%%%%%%%%%%%%%%%%%%%%%%%%%%%%
%%%%%%%%%%%%%%%%%%%%%%%%%%%%%%%%%%%%%%%%%%%%%%%%%%%%%%%%%%%%%%%%%%%%%%%%%%%
\section{Phase transition induced by varying {\Large{$\sigma$}} at low temperature near {\Large{$\sigma=2$}}}

As shown in the phase diagram (Figure 1 in the main text), varying $\sigma$ at low temperatures ($T$) leads to a transition at $\sigma=2$, as illustrated in Figure 4 of the main text. In this section, we apply the previously conjectured scaling forms of $\chi_k$ and $\xi/L$ at $\sigma=2$ to fit the critical $\sigma$ at a low temperature ($\beta=1$), similarly to fitting the critical temperature.

As mentioned previously, due to $\eta_\ell=0$ in the case of $\sigma=2$, there are certain distinctive characteristics, e.g., $\chi_k \propto 1/\ln{(L/L_0)}$, $\xi/L \propto (\ln{(L/L_0)})^{\frac{1}{2}}$, which were fitted in the last section. In what follows, we introduce the correction form and fit $\chi_k$ and $\xi/L$, respectively.

\begin{figure}[htbp!]
    \centering
    \includegraphics[width=0.7\linewidth]{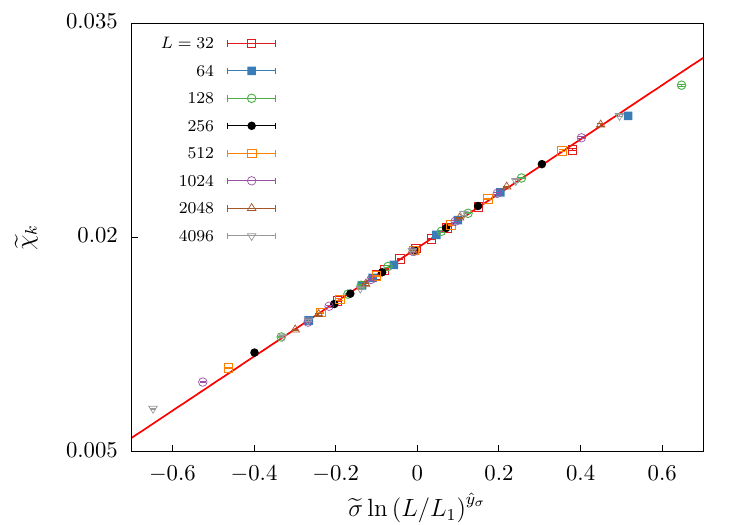}
    \caption{Data collapse according to the fitting form in Eq. \eqref{se2}: $\widetilde\chi_k = \frac{\ln{(L/L_0)}}{L^2} \chi_k - c_1L^{y_1}$. The values for each parameter are listed in Table~\ref{st2}.}
    \label{sf1}
\end{figure}

\textit{Fitting of $\chi_k$.} Near $\sigma=2$, we propose:
\begin{equation}
\chi_k=\frac{L^2}{\ln{(L/L_0)}} g(\widetilde\sigma \ln{(L/L_1)}^{\hat y_\sigma},uL^{y_u}).
\end{equation}
where $\widetilde\sigma = \frac{\sigma - \sigma_c}{\sigma_c}$, and $uL^{y_u}$ represents the irrelevant term. After a Taylor expansion, the specific fitting formula is:
\begin{equation}
\label{se2}
\chi_k=\frac{L^2}{\ln{(L/L_0)}}[a_0+b_1\widetilde\sigma \ln{(L/L_1)}^{\hat y_\sigma}+c_1L^{y_1}],
\end{equation}
where $y_1 < 0$, is the exponent for the irrelevant term. According to the fitting of $\chi_k$ at $\beta=1$ in the last section (Table~\ref{LT2}), $L_0 = e^{1.1}$. We try different values of $y_1$ during the fitting process, and the best results are shown in Table~\ref{st2}. According to the fitting results, the data points collapse onto a single curve in Fig.~\ref{sf1}.

\begin{figure}[htbp!]
    \centering
    \includegraphics[width=0.7\linewidth]{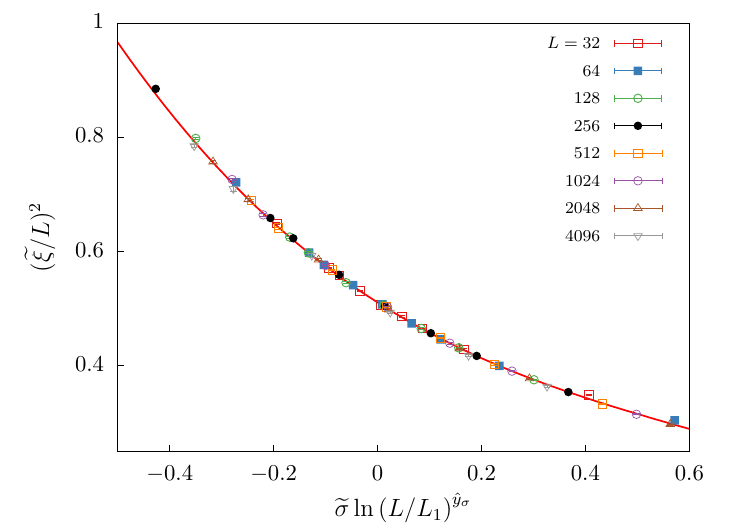}
    \caption{Data collapse according to the fitting form in Eq. \eqref{se4}: $(\widetilde\xi/L)^2 = {\ln{(L/L_0)}}^{-1} \cdot (\xi/L)^2 - c_1L^{y_1}$. The values for each parameter are listed in Table~\ref{st4}.}
    \label{sf2}
\end{figure}

\textit{Fitting of $\xi/L$.} Similarly, near $\sigma=2$, we propose:
\begin{equation}
(\xi/L)^2 = \ln{(L/L_0)} g(\widetilde\sigma \ln{(L/L_1)}^{\hat y_\sigma},uL^{y_u}).
\end{equation}
After a Taylor expansion, the specific fitting formula is:
\begin{equation}
\label{se4}
(\xi/L)^2 = \ln{(L/L_0)}(a_0 + b_1\widetilde\sigma \ln{(L/L_1)}^{\hat y_\sigma} + b_2\widetilde\sigma^2 \ln{(L/L_1)}^{2\hat y_\sigma} + b_3\widetilde\sigma^3 \ln{(L/L_1)}^{3\hat y_\sigma} + c_1L^{y_1}),
\end{equation}
where $L_0 = e^{-0.33}$, as determined by the previous fitting results (Table~\ref{LT6}). The fitting results are presented in Table~\ref{st4}. According to these results, the data points collapse onto a single curve in Fig.~\ref{sf2}.

As shown in the tables, $\sigma_c$ obtained from the fitting is always around $2$, which demonstrates the consistency of our results.

\begin{table}[!ht]
    \centering
    \caption{Fits of $\chi_k$ to \eqref{se2}, $y_1$ chosen to be -1}
    \begin{tabular}{p{2cm}p{2cm}p{2cm}p{2cm}p{2cm}p{2cm}p{2cm}p{1.2cm}}
    \hline\hline
        $L_{\rm{min}}$ & $\sigma_c$ & $L_1$ & $\hat y_\sigma$ & $a_0$ & $b_1$ & $c_1$ & $\chi^2$/DF \\ \hline
        16 & 2.006(1) & 5.4(3) & 0.91(2) & 0.01931(3) & 0.0176(7) & -0.0198(5) & 77.4/72 \\ 
        32 & 2.005(1) & 6.2(7) & 0.87(3) & 0.01926(5) & 0.019(1) & -0.018(1) & 65.6/65 \\ 
        64 & 2.007(2) & 6(1) & 0.88(5) & 0.01935(8) & 0.019(2) & -0.023(3) & 52.7/52 \\ 
        128 & 2.005(3) & 10(3) & 0.79(8) & 0.0193(1) & 0.024(4) & -0.017(8) & 42.6/43 \\ 
        256 & 2.011(6) & 3(5) & 1.0(3) & 0.0196(2) & 0.01(1) & -0.05(2) & 29.5/30 \\ \hline\hline
    \end{tabular}\label{st2}
\end{table}

\begin{table}[!ht]
    \centering
    \caption{Fits of $\xi/L$ to \eqref{se4}, $y_1$ chosen to be -1}
    % \begin{tabular}{{p{1cm}p{1cm}p{1cm}p{1cm}p{1cm}p{1cm}p{1cm}p{1cm}p{1cm}p{0.8cm}}
    \begin{tabular}{llllllllll}
    \hline\hline
        $L_{\rm{min}}$ & $\sigma_c$ & $L_1$ & $\hat y_\sigma$ & $a_0$ & $b_1$ & $b_2$ & $b_3$ & $c_1$ & $\chi^2$/DF \\ \hline 
        32 & 1.992(1) & 6.3(5) & 0.96(3) & 0.511(1) & -0.58(3) & 0.52(7) & -0.28(6) & -1.12(3) & 67.9/63 \\ 
        64 & 1.990(1) & 6(1) & 1.00(5) & 0.515(2) & -0.54(6) & 0.4(1) & -0.21(8) & -1.27(8) & 45.6/50 \\ 
        128 & 1.985(2) & 10(2) & 0.89(6) & 0.522(3) & -0.7(1) & 0.8(2) & -0.5(2) & -1.8(2) & 32.0/41 \\ 
        256 & 1.989(3) & 3(3) & 1.1(2) & 0.516(5) & -0.4(2) & 0.2(2) & -0.1(1) & -1.1(5) & 19.1/28 \\ \hline\hline
    \end{tabular}\label{st4}
\end{table}

%%%%%%%%%%%%%%%%%%%%%%%%%%%%%%%%%%%%%%%%%%%%%%%%%%%%%%%%%%%%%%%%%%%%%%%%%%%
%%%%%%%%%%%%%%%%%%%%%%%%%%%%%%%%%%%%%%%%%%%%%%%%%%%%%%%%%%%%%%%%%%%%%%%%%%%
%%%%%%%%%%%%%%%%%%%%%%%%%%%%%%%%%%%%%%%%%%%%%%%%%%%%%%%%%%%%%%%%%%%%%%%%%%%
\section{The properties of the specific heat-like quantity}
\begin{figure}[b]
    \centering
    \includegraphics[width=0.9\linewidth]{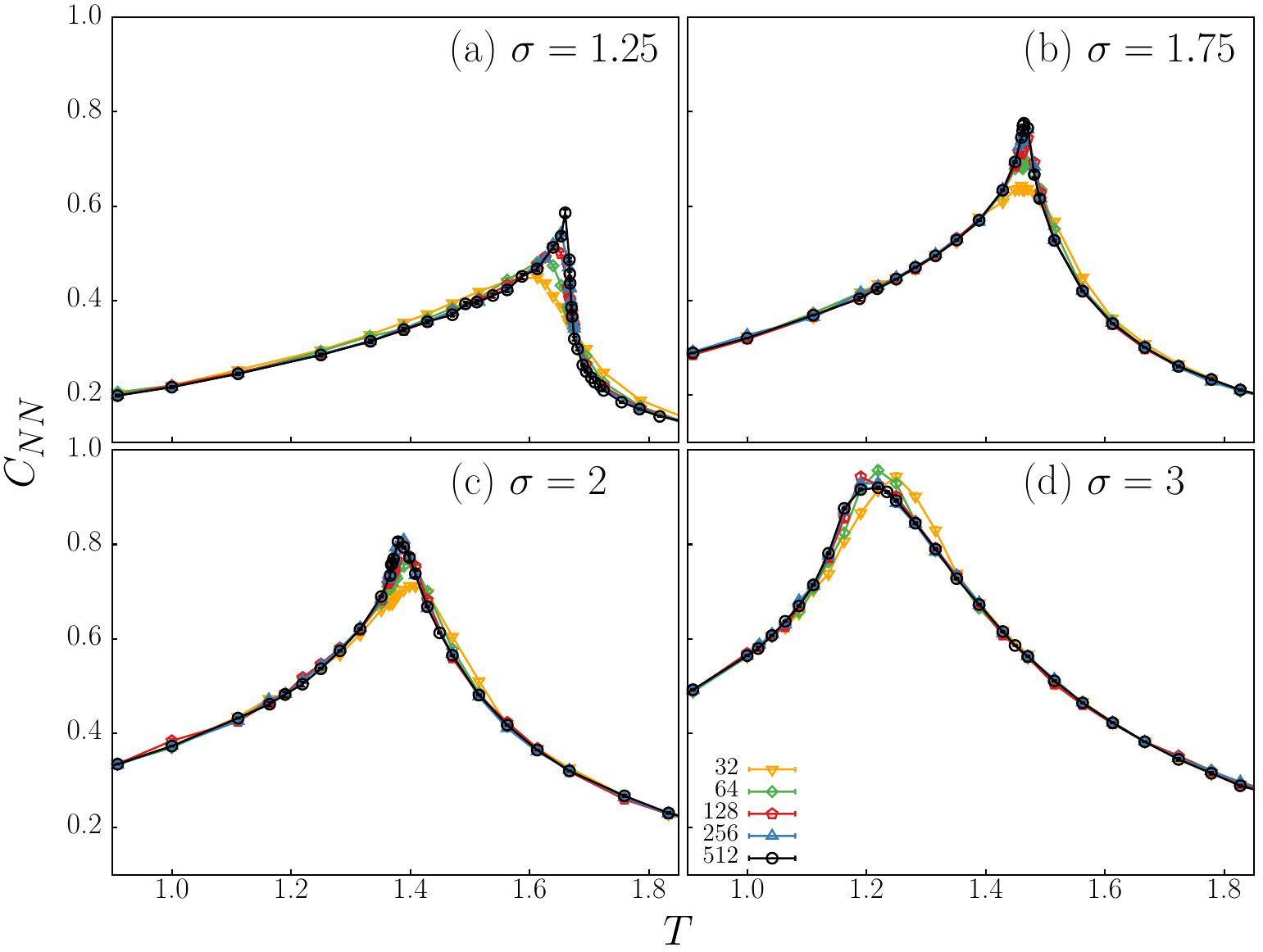}	
    \caption{The plot of $C_{NN}$ versus $T$ is shown for $\sigma = 1.25$ (a), $\sigma = 1.75$ (b), $\sigma = 2$ (c), and $\sigma = 3$ (d). The sharp peaks for $\sigma \leq 2$ indicate a second-order phase transition. In contrast, the peak for $\sigma=3$ appears smoother, indicating a BKT phase transition.}
    \label{fig:Cv}
	\end{figure}

In Fig.~\ref{fig:Cv}, we plot the specific heat-like quantity $C_{NN}$ versus temperature $T$ for $\sigma = 1.25, 1.75, 2$ and $3$. Here, the quantity is defined as 
\begin{equation}
    C_{NN} = L^2 \left( \langle \varepsilon^2\rangle - \langle \varepsilon  \rangle^2 \right),
\end{equation}
where the nearest-neighbor energy defined as $\varepsilon  = L^{-2}\sum_{\langle ij \rangle} \bm{S}_{i} \cdot \bm{S}_{j}$. Noting the computational complexity in measurements arising from long-range interactions, we only compute the energy of nearest neighbors.

We can see that the behaviors are slightly different for $\sigma \leq 2$ and $\sigma > 2$. For $\sigma = 1.25, 1.75$ and 2, the peaks are sharper than in the $\sigma = 3$ case, which characterizes a continuous phase transition. In contrast, the peak is smoother for $\sigma = 3$, indicating a BKT phase transition.
Thus, this implies that a continuous phase transition occurs for $\sigma \leq 2$, while a BKT phase transition occurs for $\sigma > 2$.

%%%%%%%%%%%%%%%%%%%%%%%%%%%%%%%%%%%%%%%%%%%%%%%%%%%%%%%%%%%%%%%%%%%%%%%%%%%
%%%%%%%%%%%%%%%%%%%%%%%%%%%%%%%%%%%%%%%%%%%%%%%%%%%%%%%%%%%%%%%%%%%%%%%%%%%
%%%%%%%%%%%%%%%%%%%%%%%%%%%%%%%%%%%%%%%%%%%%%%%%%%%%%%%%%%%%%%%%%%%%%%%%%%%